\documentclass[aps,prd,twocolumn,superscriptaddress]{revtex4}
\usepackage{epsfig,epsf}
\usepackage{amsmath}
\usepackage{amsthm}
\usepackage{amsfonts}
\usepackage{amssymb}
\usepackage{dsfont}
\usepackage{multirow}
\usepackage{appendix}
\usepackage{slashed}
\usepackage[active]{srcltx}
\usepackage{psfrag}

\begin{document}

\title{Open charm-bottom scalar tetraquarks and their strong decays }
\date{\today}
\author{S.~S.~Agaev}
\affiliation{Institute for Physical Problems, Baku State University, Az--1148 Baku,
Azerbaijan}
\author{K.~Azizi}
\affiliation{Department of Physics, Do\v{g}u\c{s} University, Acibadem-Kadik\"{o}y, 34722
Istanbul, Turkey}
\author{H.~Sundu}
\affiliation{Department of Physics, Kocaeli University, 41380 Izmit, Turkey}

\begin{abstract}
The mass and meson-current coupling of the diquark-antidiquark states with
the quantum numbers $J^{P}=0^{+}$ and quark contents $Z_{q}=[cq][\bar {b}
\bar q ]$ and $Z_{s}=[cs][\bar {b} \bar s]$ are calculated using two-point
QCD sum rule approach. In calculations the quark, gluon and mixing
condensates up to eight dimensions are taken into account. The parameters of
the scalar tetraquarks extracted from this analysis are employed to explore
the strong vertices $Z_q B_c \pi $, $Z_q B_c \eta $ and $Z_s B_c \eta $ and
compute the couplings $g_{Z_qB_c \pi }$, $g_{Z_qB_c \eta }$ and $g_{Z_sB_c
\eta }$. The strong couplings are obtained within the soft-meson
approximation of QCD light-cone sum rule method: they form, alongside with
other parameters, the basis for evaluating the widths of $Z_q \to B_c \pi$, $%
Z_q \to B_c \eta$ and $Z_s \to B_c \eta$ decays. Obtained in this work
results for the mass of the tetraquarks $Z_{q}$ and $Z_{s}$ are compared
with available predictions presented in the literature.
\end{abstract}

\maketitle

\section{Introduction}

\label{sec:Int}

During last decade the various experimental collaborations reported on
observation of hadronic states, which can not be described as the
traditional hadrons composed of two or three valence quarks. Indeed,
starting from the discovery of the $X(3872)$ state by the Belle
Collaboration \cite{Belle:2003} (see, also Ref.\ \cite{D0:2004})
measurements of various annihilation, collision and decay processes lead to
valuable experimental data on the $XYZ$ family of exotic particles.

Situation with the theoretical models, computational methods and schemes
proposed to explain observed features of the exotic states is more
complicated. One of the essential problems here is revealing the quark-gluon
structure of the exotic hadrons. Thus, in accordance with existing
theoretical models the exotic hadrons are four-quark (tetraquarks),
five-quark (pentaquarks) states, or contain as constituents valence gluons
(hybrids, glueballs). The second question is the internal quark-gluon
organization and new constituents (diquarks, antidiquarks, conventional
mesons, etc.) of the exotic hadrons, as well as a nature of the forces
binding them into compact states. Finally, one has to determine
computational methods, which can be applied to carry out qualitative
analysis these multi-quark systems. In other words, one needs to adapt to
the exotic hadrons the well known methods, which were successfully used to
explore conventional mesons and baryons, and/or to invent approaches to
solve newly emerged problems typical for the exotic states. We have only
outlined variety of problems arising when exploring the exotic hadrons. A
rather detailed information on these theoretical methods and also on
collected experimental data can be found in numerous review papers Refs.\
\cite%
{Swanson:2006st,Klempt:2007cp,Godfrey:2008nc,Voloshin:2007dx,Nielsen:2010,Faccini:2012pj, Esposito:2014rxa,Meyer:2015eta,Chen:2016qju,Lebed:2016hpi}%
, including most recent ones \cite{Chen:2016qju,Lebed:2016hpi}, and in
references therein.

Most of the observed tetraquark states belong to the class of so-called
hidden charm or bottom particles containing the $c\bar c$ or $b \bar b $
pair. But, first principles of QCD do not forbid existence of the open charm
(or bottom) or open charm-bottom tetraquarks. Experimental information
concerning the open charm tetraquarks is restricted by the observed $%
D_{s0}^{\star}(2317)$ and $D_{s1}(2460)$ mesons, which are considered as
candidates for such exotic states. These particles were explored both as the
diquark-antidiquark states and molecules built of the conventional mesons.
The only candidate to the open bottom tetraquark is $X(5568)$ state, which
is also considering as the first particle containing four valence quarks of
different flavors. But experimental status of this particle remains
controversial and unclear. Thus, the evidence for this resonance was
reported by the D0 Collaboration in Ref.\ \cite{D0:2016mwd}, later conformed
from analysis of the semileptonic decays of $B_{s}^0$ meson in Ref.\ \cite%
{D0}. At the same time, LHCb and CMS collaborations could not prove an
existence of this state on the basis of relevant experimental data [Refs.\
\cite{Aaij:2016iev,CMS:2016}]. Numerous theoretical studies of $X(5568)$ state also
suffer from contradictory conclusions ranging from conforming its parameters
measured by the D0 Collaboration till explaining the observed experimental
output by some alternative effects. Avoiding here further details, we refer
to original works addressed various aspects of the $X(5568)$ physics, and
also to the review paper devoted to the open charm and bottom mesons [Ref.\
\cite{Chen:2016spr}].

The open charm-bottom tetraquarks form the next class of the exotic
particles. It is worth noting they have not been discovered experimentally,
and to our best knowledge, there are not under consideration candidates for
these states. Nevertheless, the open charm-bottom states attracted already
interest of theorists, which performed their analysis within both the
molecule [Refs.\ \cite%
{Zhang:2009vs,Zhang:2009em,Sun:2012sy,Albuquerque:2012rq}] and
diquark-antidiquark pictures [Refs.\ \cite%
{Zouzou:1986qh,SilvestreBrac:1993ry,Chen:2013aba}] of the tetraquark model.
Thus, in Ref.\ \cite{Chen:2013aba} the authors considered the scalar and
axial-vector open charm-bottom tetraquarks and calculated their masses by
means of QCD two-point sum rules. In this article some possible decay
channels of these states are emphasized, as well.

In the present work we are going to study the scalar open charm-bottom
exotic states $Z_{q}=[cq][\bar{b}\bar{q}]$ and $Z_{s}=[cs][\bar{b}\bar{s}]$
built of the diquarks $[cq],\,[cs]$ and antidiquarks $[\bar{b}\bar{q}],\,[%
\bar{b}\bar{s}]$, where $q$ is one of the light $u$ and $d$ quarks. First we
calculate the masses and meson-current couplings of these still hypothetical
tetraquarks. To this end, we utilize QCD two-point sum rule approach, which
is one of the powerful nonperturbative methods to calculate the parameters
of the hadrons \cite{Shifman:1979}. Originally proposed to find masses,
decay constants, form factors of the conventional mesons and baryons,
it was successfully applied to analyze also exotic tetraquark states,
glueballs and hybrid $q\overline{q}g $ resonances in Refs.\ \cite{Shifman:1979,Braun:1985ah,Braun:1988kv,Balitsky:1982ps,Reinders:1985}.
The QCD two-point sum rule method remains among the fruitful computational
tools high energy physics to investigate the exotic states.

Next, we use obtained by this way parameters of the open charm-bottom
tetraquarks to explore the strong vertices $Z_{q}B_{c}\pi $, $Z_{q}B_{c}\eta
$ and $Z_{s}B_{c}\eta$ and calculate the corresponding couplings $%
g_{Z_{q}B_{c}\pi }$, $g_{Z_{q}B_{c}\eta }$ and $g_{Z_{s}B_{c}\eta }$
necessary for evaluating the widths of $Z_{q}\rightarrow B_{c}\pi $, $%
Z_{q}\rightarrow B_{c}\eta $ and $Z_{s}\rightarrow B_{c}\eta $ decays. For
these purposes, we employ QCD light-cone sum method and soft-meson
approximation suggested and elaborated in Refs.\ \cite%
{Braun:1989,Ioffe:1983ju,Braun:1995}. This method in conjunction with the
soft-meson approximation was adapted for investigation of the strong
vertices consisting of a tetraquark and two conventional mesons in Ref.\
\cite{Agaev:2016dev}. Later it was applied to calculate the decay width of
the $X(5568)$ resonance and its charmed partner state (see, Refs.\ \cite%
{Agaev:2016ijz,Agaev:2016lkl,Agaev:2016urs}). The full version of the
light-cone method was employed to analyze the strong vertices containing two
tetraquarks, as well as to compute the magnetic moment some of the
four-quark states in Refs.\ \cite{Agaev:2016srl} and \cite{Agamaliev:2016wtt}%
, respectively.

The present work is organized in the following way. In Sec.\ \ref{sec:Mass}
we calculate the masses and meson-current couplings of the scalar open
charm-bottom tetraquarks. Here we also compare our results with predictions
made in other papers. Section \ref{sec:Width} is devoted to computation of
the strong couplings corresponding to the vertices $Z_{q}B_{c}\pi $, $%
Z_{q}B_{c}\eta$ and $Z_{s}B_{c}\eta$. In this section we calculate the
widths of the decay modes $Z_{q}\rightarrow B_{c}\pi $, $Z_{q}\rightarrow
B_{c}\eta $ and $Z_{s}\rightarrow B_{c}\eta $. It contains also our brief
conclusions. We collect the spectral densities obtained in mass sum rules in the Appendix.


\section{Mass and meson-current coupling}

\label{sec:Mass}

To evaluate the masses and meson-current couplings of the
diquark-antidiquark $Z_{q}=[cq][\bar{b}\bar{q}]$ and $Z_{s}=[cs][\bar{b}\bar{%
s}]$ states we use the two-point QCD sum rules. We present explicitly
expressions necessary for computing the mass and meson-current coupling in
the case of the exotic $Z_{q}$ state. The similar formulas for the particle $%
Z_{s}$ can be obtained from a similar manner.

The scalar tetraquark state $Z_{q}=[cq][\bar{b}\bar{q}]$ can be modeled
using various interpolating currents [Ref.\ \cite{Chen:2013aba}]. To carry
out required calculations we choose the interpolating current in the form
\begin{equation}
J^{q}=q_{a}^{T}C\gamma _{5}c_{b}\left( \overline{q}_{a}\gamma _{5}C\overline{%
b}_{b}^{T} + \overline{q}_{b}\gamma _{5}C\overline{b}_{a}^{T}\right),
\label{eq:curr}
\end{equation}%
which is symmetric under exchange of the color indices $a \leftrightarrow b$%
. Here $C$ is the charge conjugation matrix. For simplicity, in what follows
we omit the superscript in the expressions.

The correlation function for the current $J^q(x)$ is given as
\begin{equation}
\Pi (p)=i\int d^{4}xe^{ip x}\langle 0|\mathcal{T}\{J^q(x)J^{q\dag
}(0)\}|0\rangle .  \label{eq:CorrF1}
\end{equation}
To derive QCD sum rule expressions for mass and meson-current coupling the
correlation function has to be calculated using both the physical and
quark-gluon degrees of freedom.

We compute the function $\Pi ^{\mathrm{Phys}}(p)$ by suggesting, that the
tetraquarks under consideration are the ground states in the relevant
hadronic channels. After saturating the correlation function with a complete set
of the $Z_q$ state and  performing in Eq.\ (\ref{eq:CorrF1}) integral over $x$ ,
we get the required expression for $\Pi ^{\mathrm{Phys}}(p)$
\begin{equation*}
\Pi ^{\mathrm{Phys}}(p)=\frac{\langle 0|J^q|Z_{q}(p)\rangle \langle
Z_{q}(p)|J^{q\dag }|0\rangle }{m_{Z_q}^{2}-p^{2}}+...
\end{equation*}%
where $m_{Z_q}$ is the mass of the $Z_{q}$ state, and dots stand for
contributions of the higher resonances and continuum states. We define the
meson-current coupling by the equality%
\begin{equation*}
\langle 0|J^q|Z_{q}(p)\rangle =f_{Z_q}m_{Z_q}.
\end{equation*}%
Then in terms of $m_{Z_q}$ and $f_{Z_q}$ the correlation function takes the
simple form
\begin{equation}
\Pi ^{\mathrm{Phys}}(p)=\frac{m_{Z_q}^{2}f_{Z_q}^{2}}{m_{Z_q}^{2}-p^{2}}+\ldots
\label{eq:CorM}
\end{equation}%
It contains only one term, which is proportional to the identity matrix,
and, therefore, can be replaced by the invariant function $\Pi ^{\mathrm{Phys%
}}(p^2)$. The Borel transformation applied to these invariant function yields%
\begin{equation}
\mathcal{B}_{p^{2}}\Pi ^{\mathrm{Phys}%
}(p^2)=m_{Z_q}^{2}f_{Z_q}^{2}e^{-m_{Z_q}^{2}/M^{2}}+\ldots  \label{eq:CorBor}
\end{equation}

In order to obtain the function $\Pi(p)$ using the quark-gluon degrees of
freedom, i.e. by employing the light and heavy propagators, we substitute
the interpolating current given by Eq.\ (\ref{eq:curr}) into Eq.\ (\ref%
{eq:CorrF1}), and contract the relevant quark fields. As a result, for $\Pi
^{\mathrm{QCD}}(p)$ we get:
\begin{eqnarray}
&&\Pi ^{\mathrm{QCD}}(p)=i\int d^{4}xe^{ipx}\left\{ \mathrm{Tr}\left[ \gamma
_{5}\widetilde{S}_{b}^{b^{\prime }b}(-x)\gamma _{5}S_{q}^{aa^{\prime }}(-x)%
\right] \right.  \notag \\
&&\times \mathrm{Tr}\left[ \gamma _{5}\widetilde{S}_{q}^{aa^{\prime
}}(x)\gamma _{5}S_{c}^{bb^{\prime }}(x)\right] +\mathrm{Tr}\left[ \gamma _{5}%
\widetilde{S}_{b}^{a^{\prime }b}(-x)\right.  \notag \\
&&\times \left. \gamma _{5}S_{q}^{b^{\prime }a}(-x)\right] \mathrm{Tr}\left[
\gamma _{5}\widetilde{S}_{q}^{aa^{\prime }}(x)\gamma _{5}S_{c}^{bb^{\prime
}}(x)\right]  \notag \\
&&+\mathrm{Tr}\left[ \gamma _{5}\widetilde{S}_{b}^{b^{\prime }a}(-x)\gamma
_{5}S_{q}^{a^{\prime }b}(-x)\right] \mathrm{Tr}\left[ \gamma _{5}\widetilde{S%
}_{q}^{aa^{\prime }}(x)\gamma _{5}S_{c}^{bb^{\prime }}(x)\right]  \notag \\
&&\left. +\mathrm{Tr}\left[ \gamma _{5}\widetilde{S}_{b}^{a^{\prime
}a}(-x)\gamma _{5}S_{q}^{b^{\prime }b}(-x)\right] \mathrm{Tr}\left[ \gamma
_{5}\widetilde{S}_{q}^{aa^{\prime }}(x)\gamma _{5}S_{c}^{bb^{\prime }}(x)%
\right] \right\} ,  \notag \\
&&{}  \label{eq:CorrF2}
\end{eqnarray}%
where we employ the notation
\begin{equation}
\widetilde{S}_{q(b)}^{ab}(x)=CS_{q(b)}^{Tab}(x)C,  \label{eq:Not}
\end{equation}%
with $S_{q}(x)$ and $S_{b}(x)$ being the $q$- and $b$-quark propagators,
respectively.

We continue by invoking into analysis the well known expressions of the
light and heavy quark propagators. For our purposes it is convenient to use
the $x$-space expression of the light quark propagators, whereas for the
heavy quarks we utilize their propagators given in the momentum space. Thus,
for the light quarks we have:
\begin{eqnarray}
&&S_{q}^{ab}(x)=i\delta _{ab}\frac{\slashed x}{2\pi ^{2}x^{4}}-\delta _{ab}%
\frac{m_{q}}{4\pi ^{2}x^{2}}-\delta _{ab}\frac{\langle \overline{q}q\rangle
}{12}  \notag \\
&&+i\delta _{ab}\frac{\slashed xm_{q}\langle \overline{q}q\rangle }{48}%
-\delta _{ab}\frac{x^{2}}{192}\langle \overline{q}g_{}\sigma Gq\rangle
+i\delta _{ab}\frac{x^{2}\slashed xm_{q}}{1152}\langle \overline{q}%
g_{}\sigma Gq\rangle  \notag \\
&&-i\frac{g_{}G_{ab}^{\alpha \beta }}{32\pi ^{2}x^{2}}\left[ \slashed x{%
\sigma _{\alpha \beta }+\sigma _{\alpha \beta }}\slashed x\right] -i\delta
_{ab}\frac{x^{2}\slashed xg_{}^{2}\langle \overline{q}q\rangle ^{2}}{7776}
\notag \\
&&-\delta _{ab}\frac{x^{4}\langle \overline{q}q\rangle \langle
g_{}^{2}G^{2}\rangle }{27648}+\ldots  \label{eq:qprop}
\end{eqnarray}%
For the heavy $Q=b,\ c$ quark propagator $S_{Q}^{ab}(x)$ we utilize the
expression from Ref.\ \cite{Reinders:1984sr}.
\begin{eqnarray}
&&S_{Q}^{ab}(x)=i\int \frac{d^{4}k}{(2\pi )^{4}}e^{-ikx}\Bigg \{\frac{\delta
_{ab}\left( {\slashed k}+m_{Q}\right) }{k^{2}-m_{Q}^{2}}  \notag \\
&&-\frac{g_{}G_{ab}^{\alpha \beta }}{4}\frac{\sigma _{\alpha \beta }\left( {%
\slashed k}+m_{Q}\right) +\left( {\slashed k}+m_{Q}\right) \sigma _{\alpha
\beta }}{(k^{2}-m_{Q}^{2})^{2}}  \notag \\
&&+\frac{g_{}^{2}G^{2}}{12}\delta _{ab}m_{Q}\frac{k^{2}+m_{Q}{\slashed k}}{%
(k^{2}-m_{Q}^{2})^{4}}+\frac{g_{}^{3}G^{3}}{48}\delta _{ab}\frac{\left( {%
\slashed k}+m_{Q}\right) }{(k^{2}-m_{Q}^{2})^{6}}  \notag \\
&&\times \left[ {\slashed k}\left( k^{2}-3m_{Q}^{2}\right) +2m_{Q}\left(
2k^{2}-m_{Q}^{2}\right) \right] \left( {\slashed k}+m_{Q}\right) +\ldots %
\Bigg \}.  \notag \\
&&{}  \label{eq:Qprop}
\end{eqnarray}%
In Eqs.\ (\ref{eq:qprop}) and (\ref{eq:Qprop}) the standard notations
\begin{eqnarray}
&&G_{ab}^{\alpha \beta }=G_{A}^{\alpha \beta
}t_{ab}^{A},\,\,~~G^{2}=G_{\alpha \beta }^{A}G_{\alpha \beta }^{A},  \notag
\\
&&G^{3}=\,\,f^{ABC}G_{\mu \nu }^{A}G_{\nu \delta }^{B}G_{\delta \mu }^{C},
\end{eqnarray}%
are introduced. Here $a,\,b=1,2,3$ and $A,B,C=1,\,2\,\ldots 8$ are the color
indices, and $t^{A}=\lambda ^{A}/2$ with $\lambda ^{A}$ being the Gell-Mann
matrices. In the nonperturbative terms the gluon field strength tensor $%
G_{\alpha \beta }^{A}\equiv G_{\alpha \beta }^{A}(0)$ is fixed at $x=0.$
\begin{table}[tbp]
\begin{tabular}{|c|c|}
\hline\hline
Parameters & Values \\ \hline\hline
$m_{B_c}$ & $(6275.1 \pm 1.0) ~\mathrm{MeV}$ \\
$f_{B_c}$ & $(528 \pm 19)~\mathrm{MeV}$ \\
$m_{\eta}$ & $(547.862 \pm 0.017) ~\mathrm{MeV}$ \\
$m_{\pi}$ & $(134.9766 \pm 0.0006) ~\mathrm{MeV}$ \\
$f_{\pi}$ & $0.131~\mathrm{GeV}$ \\
$m_{b}$ & $4.18^{+0.04}_{-0.03}~\mathrm{GeV}$ \\
$m_{c}$ & $(1.27 \pm 0.03)~\mathrm{GeV}$ \\
$m_{s} $ & $96^{+8}_{-4}~\mathrm{MeV} $ \\
$\langle \bar{q}q \rangle $ & $(-0.24\pm 0.01)^3$ $\mathrm{GeV}^3$ \\
$\langle \bar{s}s \rangle $ & $0.8\ \langle \bar{q}q \rangle$ \\
$m_{0}^2 $ & $(0.8\pm0.1)$ $\mathrm{GeV}^2$ \\
$\langle \overline{q}g_{}\sigma Gq\rangle$ & $m_{0}^2\langle \bar{q}q
\rangle $ \\
$\langle \overline{s}g_{}\sigma Gs\rangle$ & $m_{0}^2\langle \bar{s}s
\rangle $ \\
$\langle\frac{\alpha_sG^2}{\pi}\rangle $ & $(0.012\pm0.004)$ $~\mathrm{GeV}%
^4 $ \\
$\langle g_{}^3G^3\rangle $ & $(0.57\pm0.29)$ $~\mathrm{GeV}^6 $ \\
\hline\hline
\end{tabular}%
\caption{Input parameters.}
\label{tab:Param}
\end{table}

Strictly speaking, the QCD sum rule expressions are derived after fixing the
same Lorentz structures in the both physical and theoretical expressions of
the correlation function. In the case of the scalar particles, as we have
just noted, the only Lorentz structure in these expressions is $\sim \mathrm{%
I}$. Hence, there is only one invariant function $\Pi ^{\mathrm{QCD}}(p^{2})$
in theoretical side of the sum rule, which can be represented as the
dispersion integral
\begin{equation}
\Pi ^{\mathrm{QCD}}(p^{2})=\int_{\mathcal{M}^2}^{\infty }\frac{\rho ^{%
\mathrm{QCD}}(s)ds}{s-p^{2}}+...,
\end{equation}%
where $\mathcal{M}=m_b+m_c+2m_q$, and $\rho ^{\mathrm{QCD}}(s)$ is the corresponding spectral density.

The spectral density $\rho ^{\mathrm{QCD}}(s)$ is the key ingredient of the
sum rule calculations. The technical methods for calculation of the spectral
density in the case of the tetraquark states are well known and presented in
rather clear form, for example, in Refs.\ \cite{Agaev:2016dev,Agaev:2016mjb}.
Therefore, here we omit details of calculations and move the final explicit expressions  obtained  for $\rho ^{\mathrm{QCD}}(s)$ corresponding to $ Z_q $ state to  the Appendix. Let us note
only that the spectral density is computed by taking into account
condensates up to dimension eight: it depends on
the quark, gluon $\langle \overline{q}q\rangle$, $\langle
g_{}^{2}G^{2}\rangle$, $\langle g_{}^{3}G^{3}\rangle$ and mixed $\langle
\overline{q}g_{}\sigma Gq\rangle$ condensates, and ones due to their
products.

Applying the Borel transformation on the variable $p^{2}$ to the invariant
function $\Pi ^{\mathrm{QCD}}(p^{2})$, equating the obtained expression with
$\mathcal{B}_{p^{2}}\Pi ^{\mathrm{Phys}}(p)$, and subtracting the
contribution arising from higher resonances and continuum states, we find
the final sum rules. Thus, the sum rule for the mass of the $Z_{q}$ state
reads
\begin{equation}
m_{Z_q}^{2}=\frac{\int_{\mathcal{M}^2}^{s_{0}}ds \rho ^{\mathrm{QCD}%
}(s)se^{-s/M^{2}}}{\int_{\mathcal{M}^2}^{s_{0}}ds\rho ^{\mathrm{QCD}%
}(s)e^{-s/M^{2}}}.  \label{eq:srmass}
\end{equation}%
The meson-current coupling $f_{Z_q}$ is given by the sum rule:
\begin{equation}
f_{Z_q}^{2}m_{Z_q}^{2}e^{-m_{Z_q}^{2}/M^{2}}=\int_{\mathcal{M}^2}^{s_{0}}ds%
\rho ^{\mathrm{QCD}}(s)e^{-s/M^{2}}.  \label{eq:srcoupling}
\end{equation}%
In Eqs.\ (\ref{eq:srmass}) and (\ref{eq:srcoupling}) by $s_0$ we denote the
threshold parameter, that dissects the contribution of the ground state from
one due to the higher resonances and continuum. Here we should remark that in the present work we calculate the meson-current couplings $ f_{Z_q} $ and $ f_{Z_s} $ for the first time: they are main input parameters for calculation of the strong coupling constants considered in the next section and were not analyzed in Ref. \ \cite{Chen:2013aba}.

The sum rules contain parameters numerical values of which should be
specified. We collect the required information in Table\ \ref{tab:Param}.
For the vacuum expectation value of the gluon field $\sim g_{}^{3}G^3$ we
employ the result reported in Ref.\ \cite{Narison:2015nxh}. The remaining
quark and gluon condensates are well known, and we utilize their standard
values. The Table\ \ref{tab:Param} contains also $B_c$, $\eta$ and $\pi$
meson masses and decay constants, which will serve as input parameters for
computing of the strong couplings and decay widths in the next section
(see, Ref.\ \cite{Olive:2016xmw}).

The QCD sum rules depend on the continuum threshold $s_{0}$ and Borel
variable $M^{2}$. To extract reliable information from the sum rules we have
to choose such regions for $s_0$ and $M^2$, where the physical quantities
under question demonstrate minimal sensitivity on them.
It is worth emphasizing, that namely these
two parameters are the main sources of uncertainties in QCD sum rule
predictions.

According to the method used, the window for the Borel parameter has to provide the convergence of the series of
operator product expansion (OPE), and suppression of the higher resonance and
continuum contributions to the sum rule. The convergence of OPE, i.e., the exceeding of the perturbative part to the nonperturbative contributions and reducing  the contribution with increasing the dimension of the nonperturbative operators are easily achieved for the exotic states like the standard hadrons. However, in the exotic channels the pole contribution to the mass sum rules remains mainly under $ 50\% $ of the total integral. But, as we will see in the next section, in the case of strong couplings of the exotic states with conventional hadrons the pole contribution exceeds $ 70\% $  of the whole result. To find the lower boundary for $ M^2 $ we demand convergence of the OPE and exceeding of the perturbative part over the nonperturbative contribution. The upper limit for this parameter is extracted by requiring largest possible pole contribution. As a result, for $ M^2 $, in the mass and meson-current calculations, we fix the following range
\begin{equation}
6.5\ \mathrm{GeV}^{2}\leq M^{2}\leq 7.5\ \mathrm{GeV}^{2}.
\end{equation}%
The choice of the continuum threshold $s_0$ depends on the energy of the
first excited state and can be extracted from analysis of the pole/total
ratio. This criterium enables us to determine the range of $s_0$ as
\begin{equation}
55\ \mathrm{GeV}^{2}\leq s_0\leq 57\ \mathrm{GeV}^{2}.
\end{equation}
To see how the OPE converges and how large is the pole contribution some plots are in order. We compare the perturbative and nonperturbative contributions to the mass sum rule by varying $ M^2 $ at fixed average value of $ s_0 $, and by varying $ s_0 $ at fixed average $ M^2 $ in the left and right panels of Fig. \ref{fig:pert_nonpert}, respectively. The contributions of different  nonperturbative operators with respect to $M^2$ at average value of the continuum threshold and the same quantity with respect to $ s_0 $ at average value of $M^2$ are presented in the left and right panels  of Fig. \ref{fig:nonpert}, respectively. The pole/total contribution that is shown by PC also on $M^2$ and $ s_0 $ are depicted in Fig. \ref{fig:PC}. 
\begin{widetext}

\begin{figure}[h!]
\begin{center}
\includegraphics[totalheight=6cm,width=8cm]{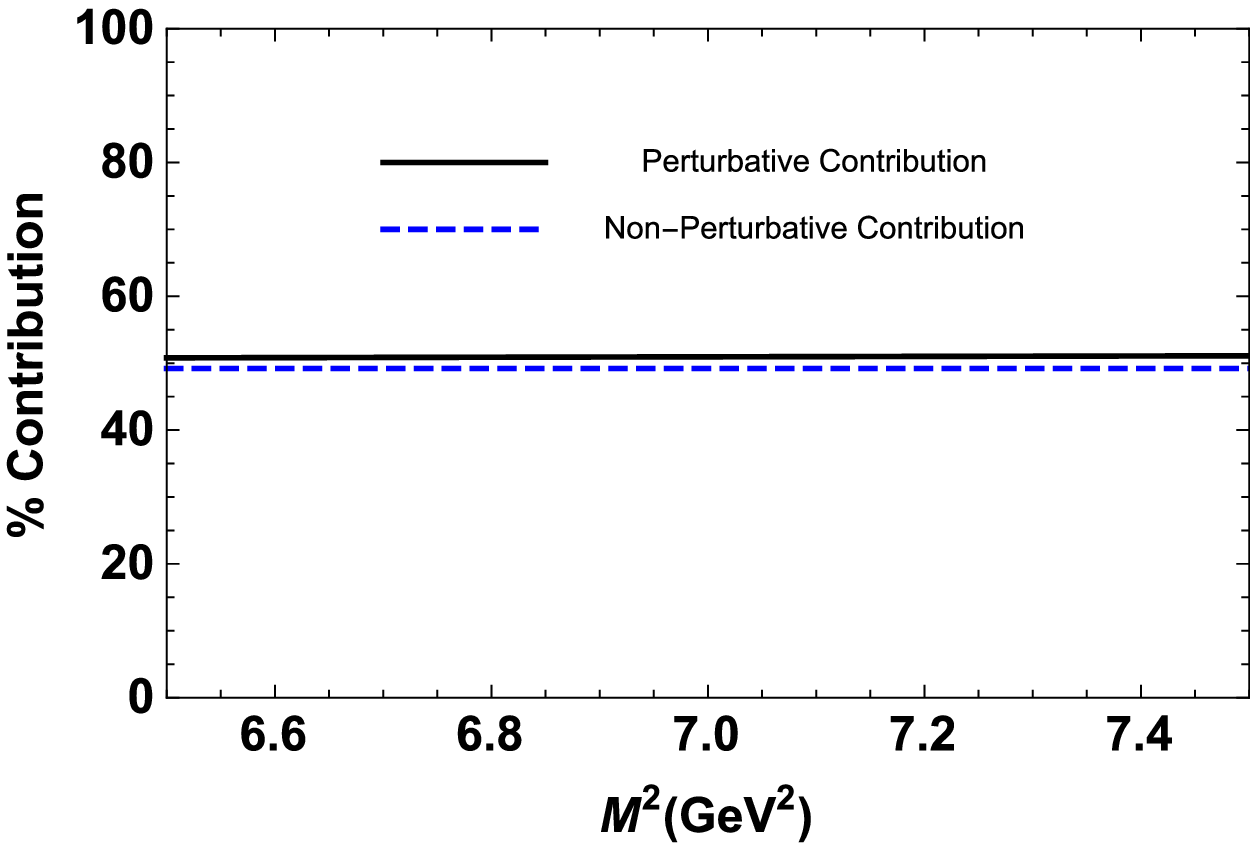}\,\,
\includegraphics[totalheight=6cm,width=8cm]{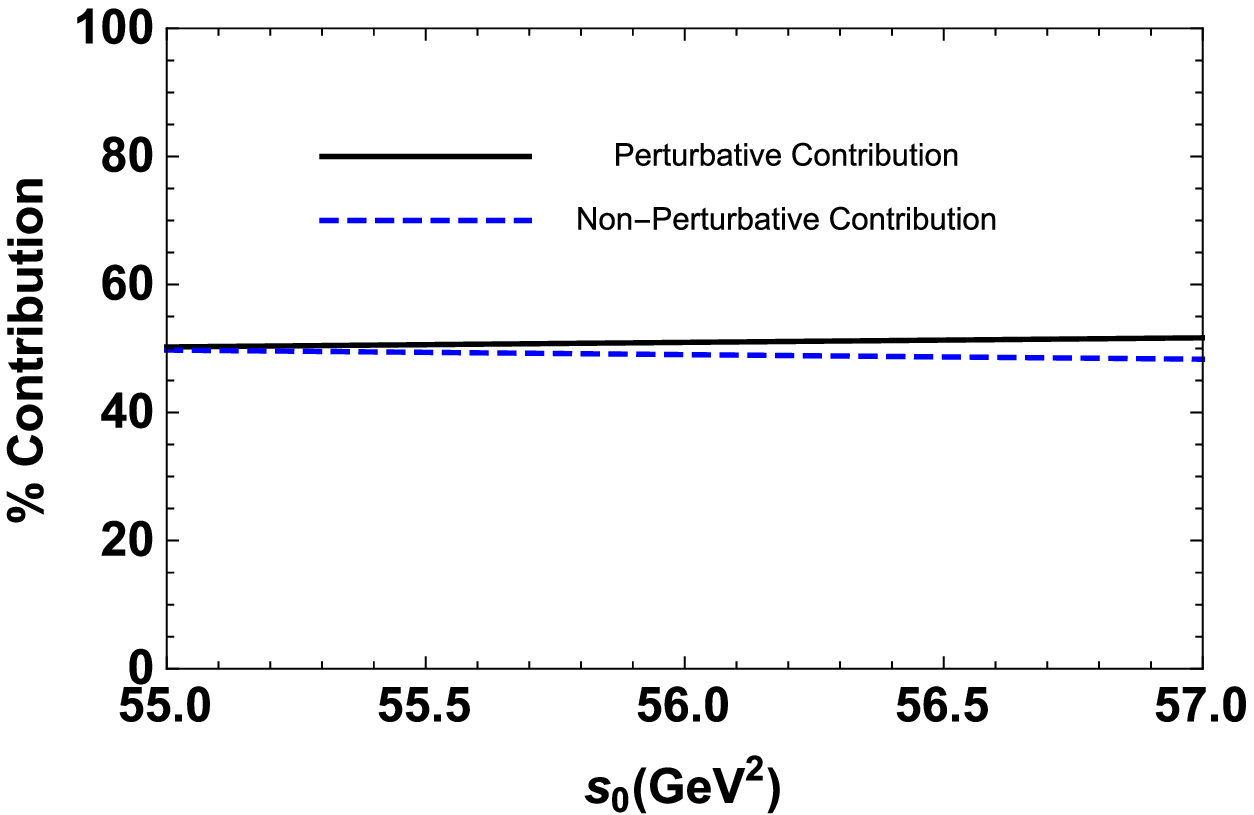}
\end{center}
\caption{\textbf{Left:} Comparison of  the perturbative and nonperturbative contributions to the mass sum rule of $ Z_q $ with respect to  $M^2$ at average value of $ s_0 $. \textbf{Right:}
 The same as left panel but in terms of $ s_0 $ at average value of the   Borel
parameter $M^2$. } \label{fig:pert_nonpert}
\end{figure}
\begin{figure}[h!]
\begin{center}
\includegraphics[totalheight=6cm,width=8cm]{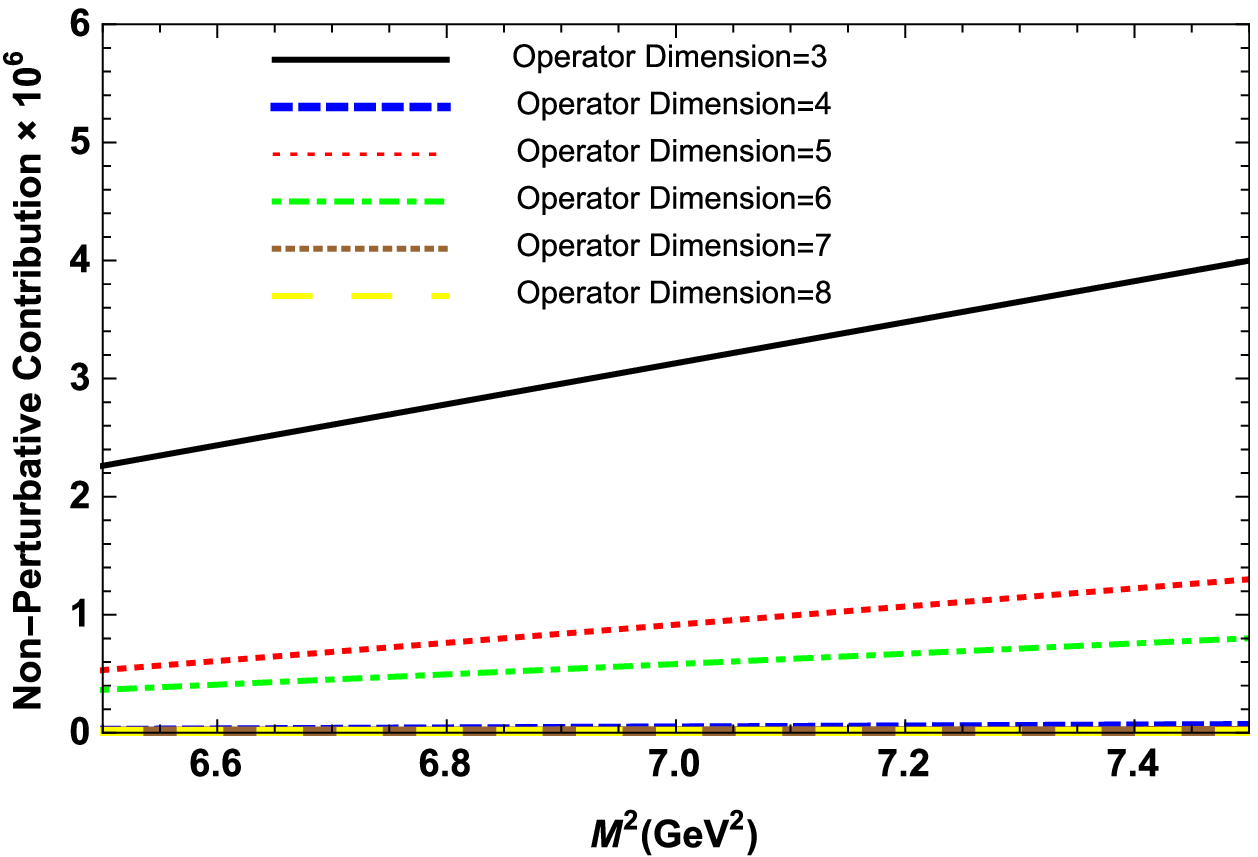}\,\,
\includegraphics[totalheight=6cm,width=8cm]{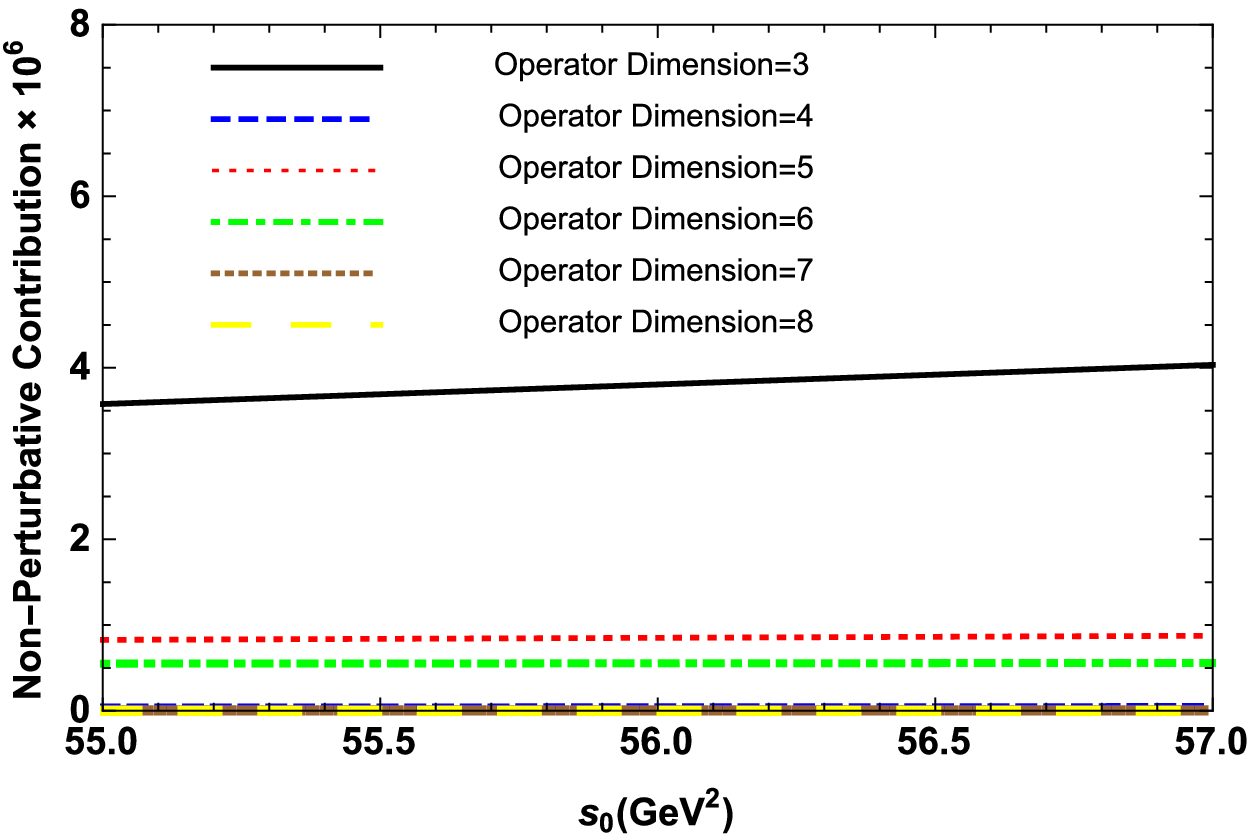}
\end{center}
\caption{\textbf{Left:} Contribution of different  nonperturbative operators to the mass sum rule of $ Z_q $ with respect to  $M^2$ at average value of $ s_0 $. \textbf{Right:}
 The same as left panel but in terms of $ s_0 $ at average value of the   Borel
parameter $M^2$. } \label{fig:nonpert}
\end{figure}
\begin{figure}[h!]
\begin{center}
\includegraphics[totalheight=6cm,width=8cm]{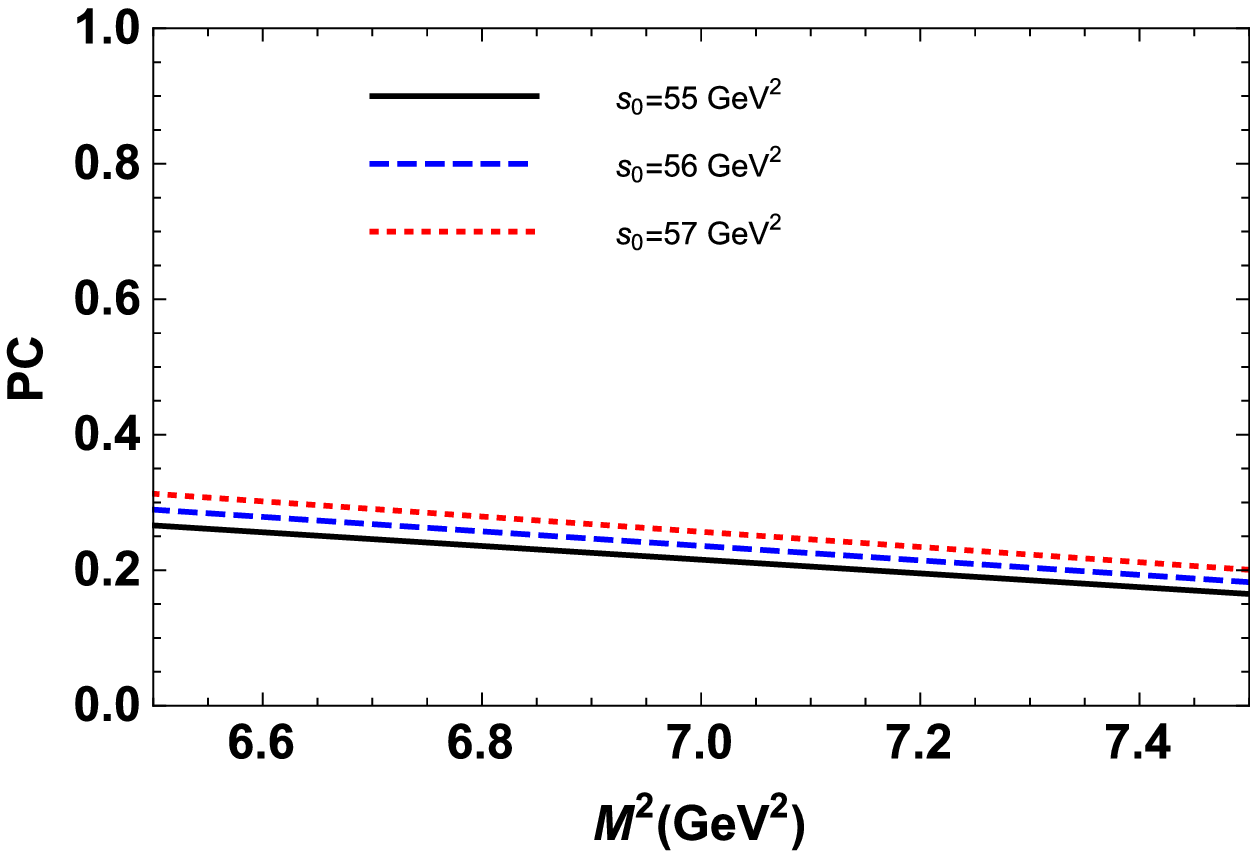}\,\,
\includegraphics[totalheight=6cm,width=8cm]{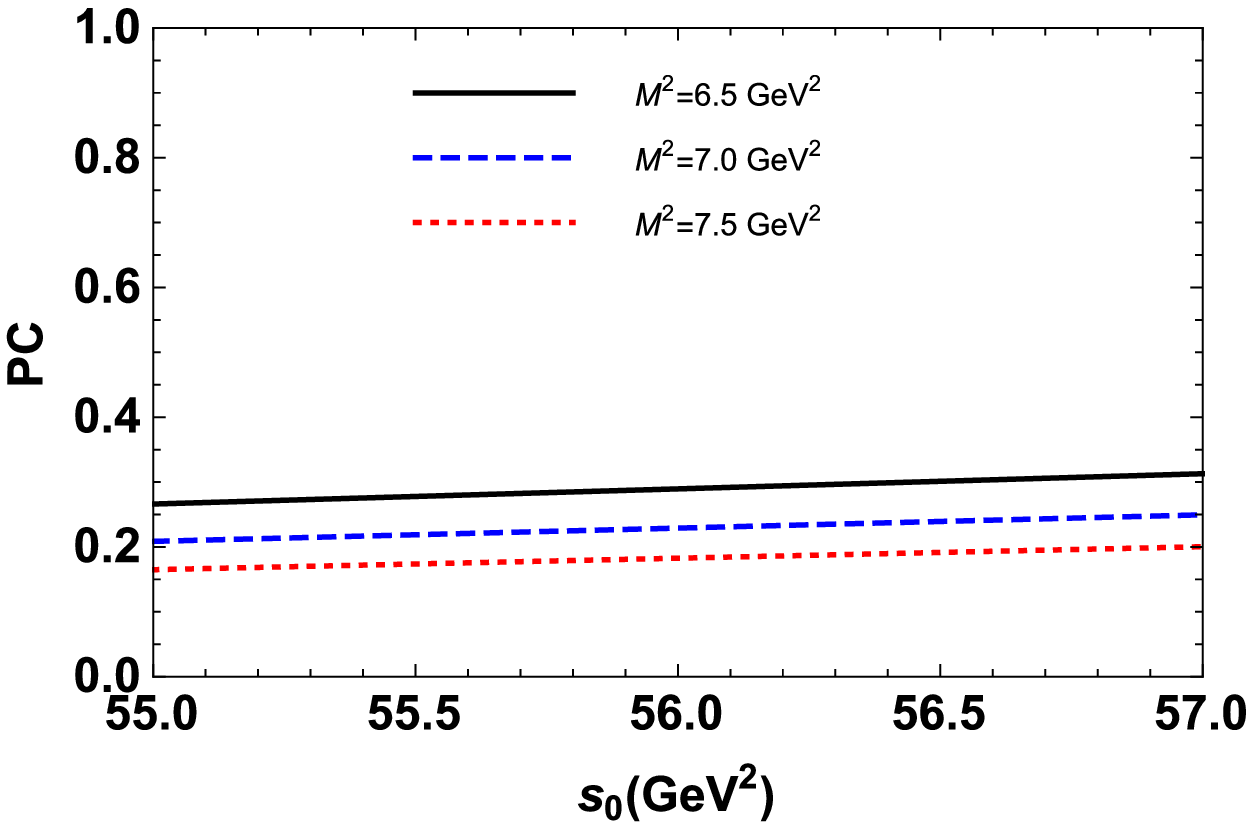}
\end{center}
\caption{\textbf{Left:} Pole/total contribution for mass sum rule of  $ Z_q $ with respect to  $M^2$ at different fixed values of $ s_0 $. \textbf{Right:}
 The same as left panel but in terms of $ s_0 $ at different fixed values of the   Borel
parameter $M^2$. } \label{fig:PC}
\end{figure}
\begin{figure}[h!]
\begin{center}
\includegraphics[totalheight=6cm,width=8cm]{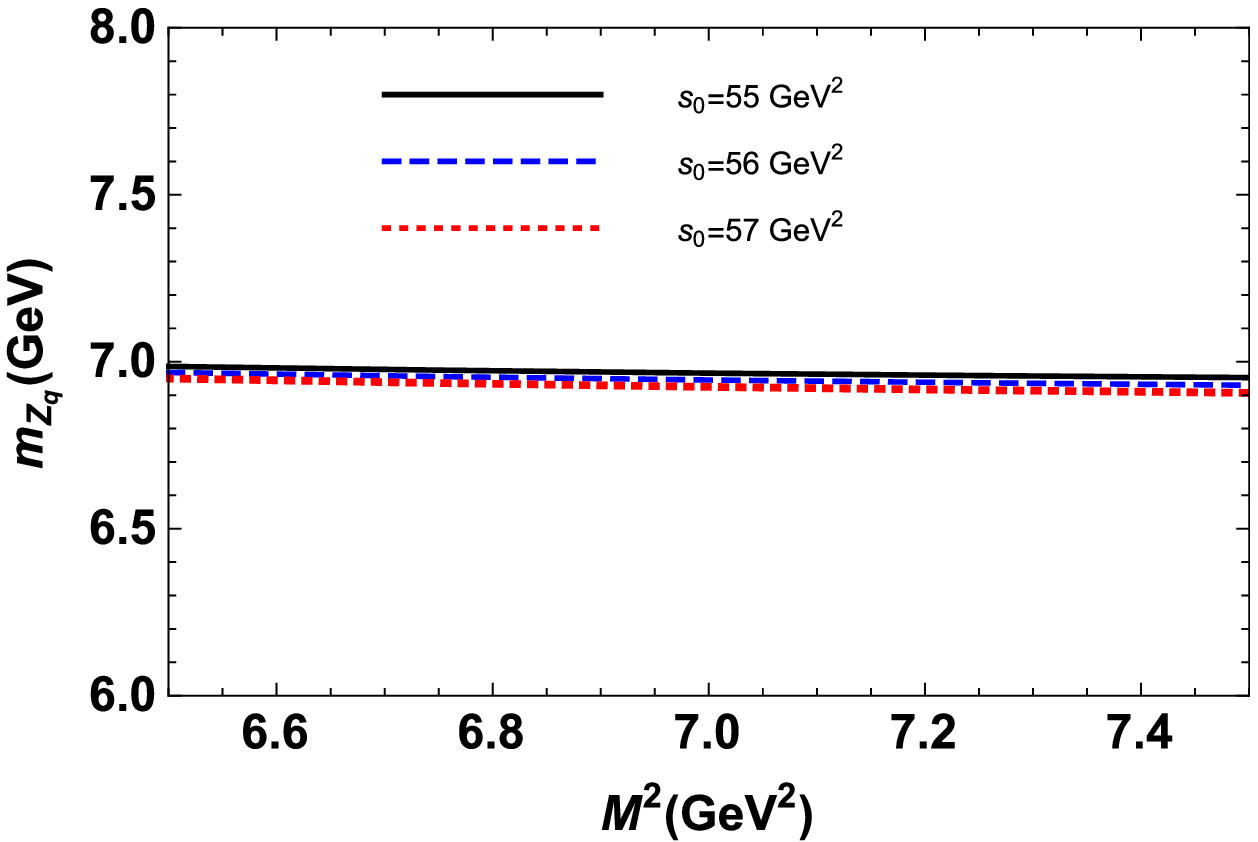}
\includegraphics[totalheight=6cm,width=8cm]{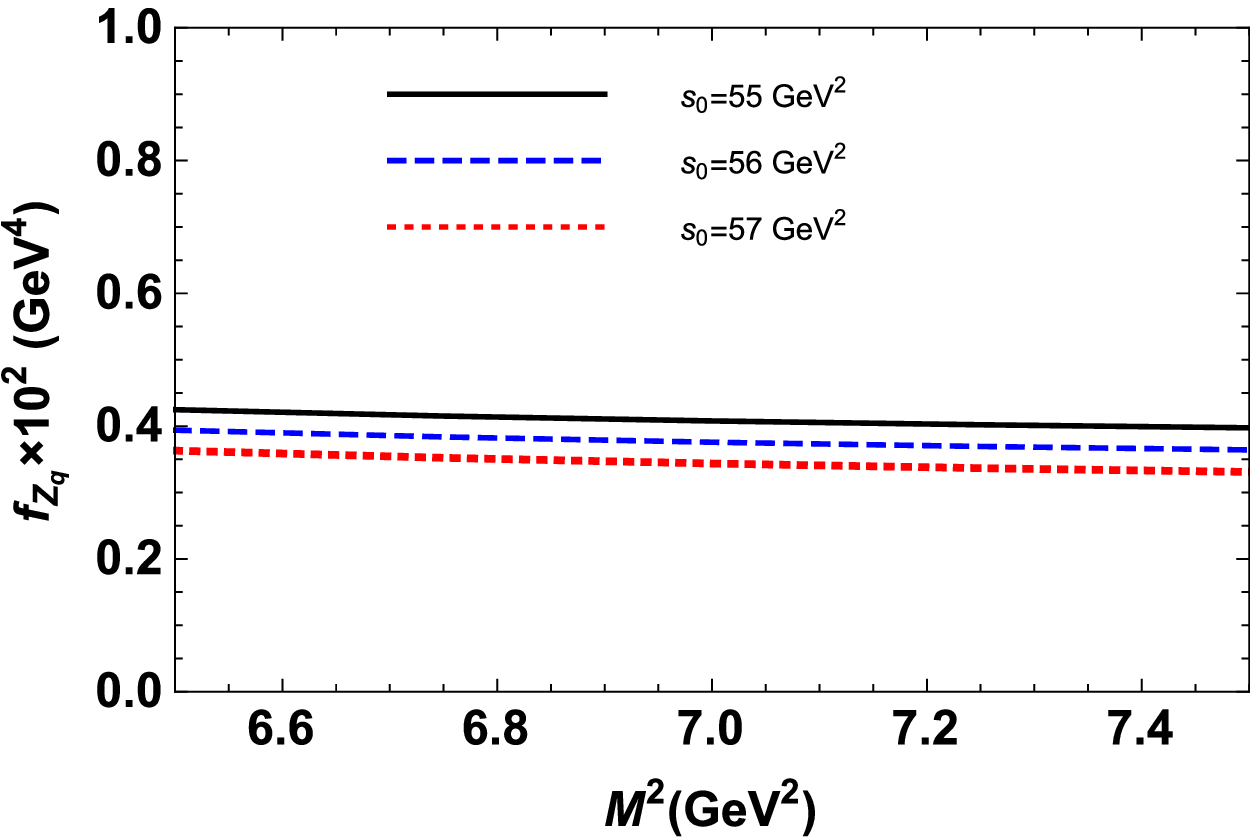}
\end{center}
\caption{\textbf{Left:} The mass of the $Z_q$ state as a function of the Borel parameter $
M^2$ at various values of $s_0$. \textbf{Right:}
 The meson-current coupling $f_{Z_q}$ as a function of the Borel
parameter $M^2$ at different values of $s_0$.} \label{fig:MassresiduZq}
\end{figure}
\begin{figure}[h!]
\begin{center}
\includegraphics[totalheight=6cm,width=8cm]{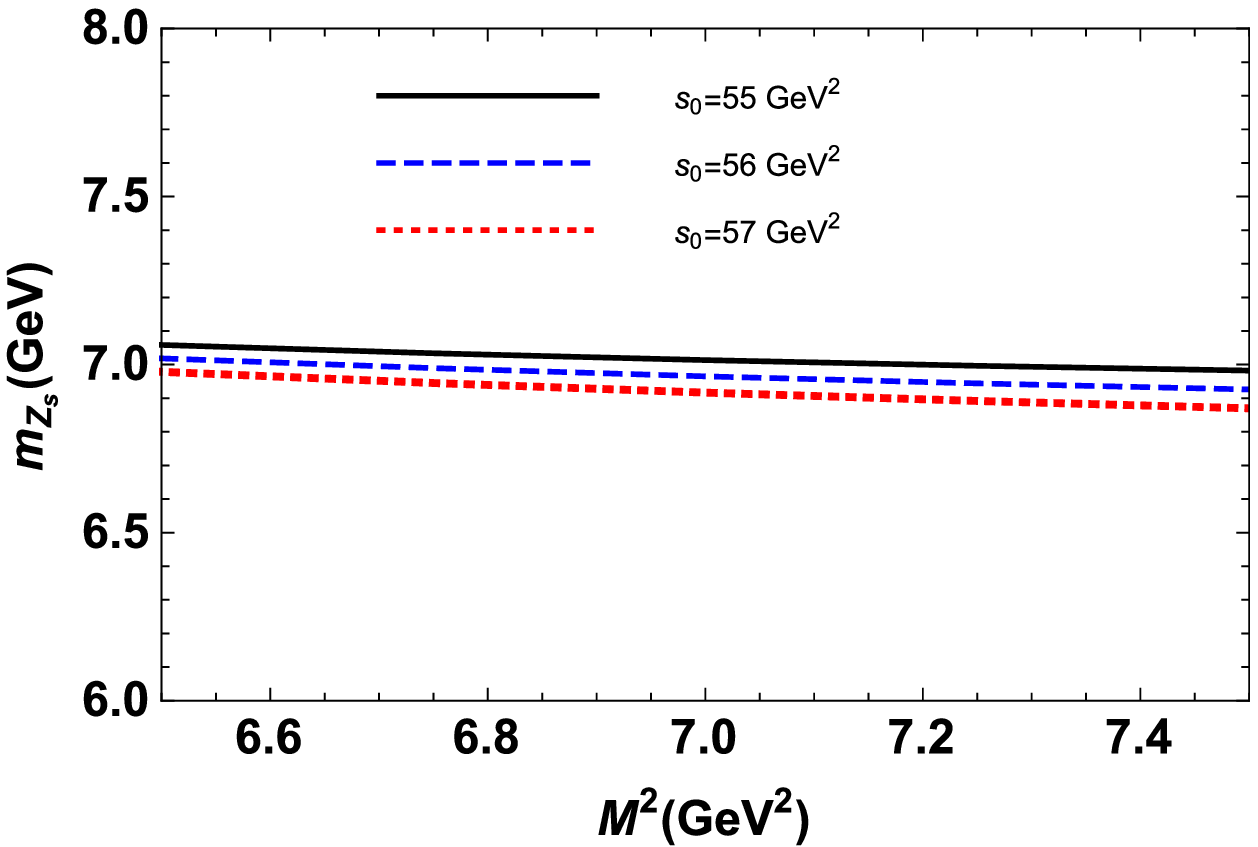}
\includegraphics[totalheight=6cm,width=8cm]{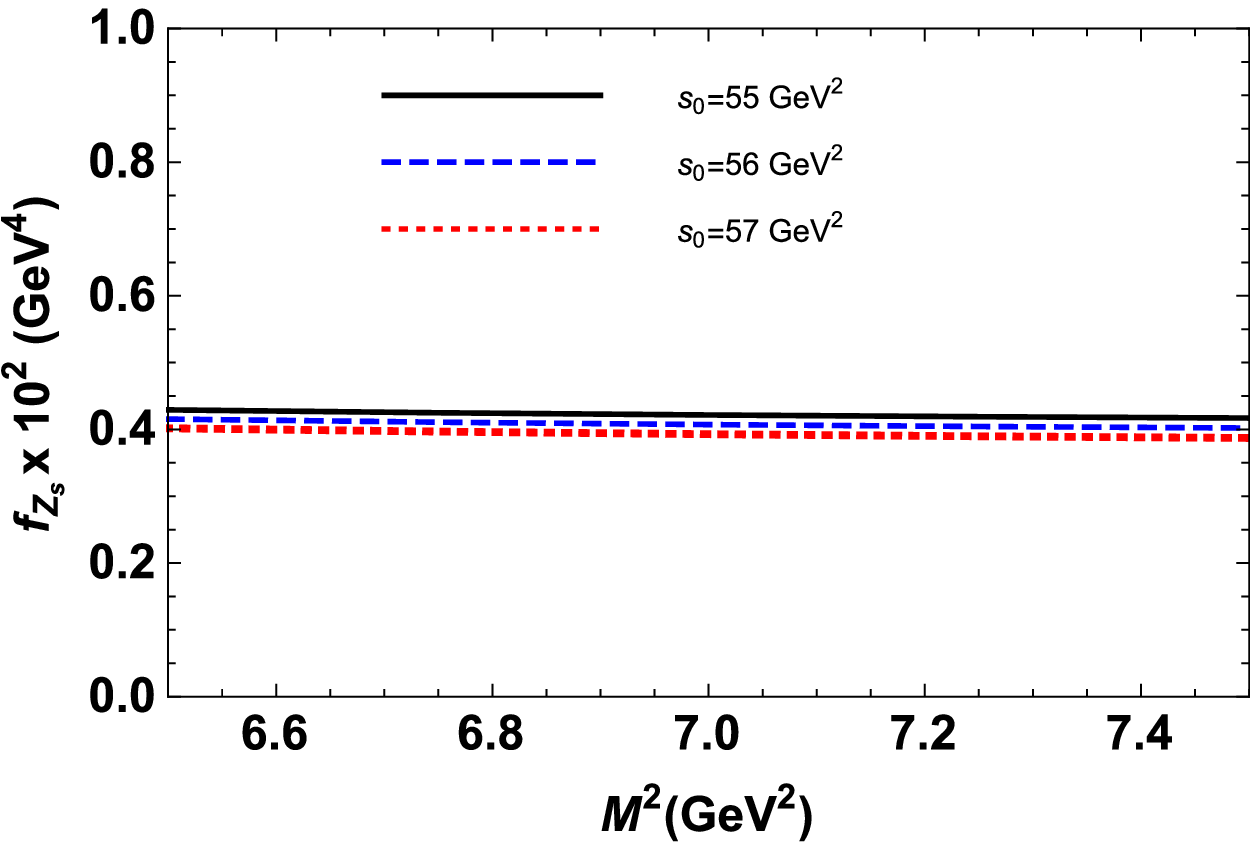}
\end{center}
\caption{\textbf{Left:} The mass of the $Z_s$ state as a function of the Borel parameter $
M^2$ at various values of $s_0$. \textbf{Right:}
 The meson-current coupling $f_{Z_s}$ as a function of the Borel
parameter $M^2$ at different values of $s_0$.} \label{fig:MassresiduZs}
\end{figure}

\end{widetext}
From these figures we see that inside of the working windows for $M^2$  and $s_0$ , the mass sum rule demonstrates a good  convergence and the perturbative part constitutes the main part of the total integral. We reach the PC contribution in the range $  (16-31)\%$ for different values of $M^2$  and $s_0$ in their working regions. We shall also remark that the  working regions for the Borel parameter and continuum threshold obtained for $ Z_q $ state are roughly  the same for $ Z_s $ state and the $ SU(3) $ flavor violation is negligible. Similar results for the convergence of OPE and pole contribution in  $ Z_q $ channel are obtained for $ Z_s $ state  as well and  the presence of $ s $ quark dose not change the situations in the figures   \ref{fig:pert_nonpert}- \ref{fig:PC}, considerably.

The results obtained for the mass and meson-current coupling of the $Z_q$ and $Z_s$
state are plotted in Figs.\ \ref{fig:MassresiduZq} and \ref{fig:MassresiduZs}, and
demonstrate mild dependence on $s_0$ and $M^2$.
Our results for the masses and meson-current couplings of the $Z_q$ and $Z_s$
states are collected in Table\ \ref{tab:Results1}. Here under $Z_q$ we imply
both the $Z_u$ and $Z_d$ states, which in the exact isospin symmetry
accepted in this work have identical physical parameters.

The masses of the scalar diquark-antidiquark states with the same contents
were calculated in Ref.\ \cite{Chen:2013aba}, as well. The authors used QCD
two-point sum rule approach, and for the masses of the $Z_s$ and $Z_q$
states found:
\begin{equation}
m_{Z_s}=7.16 \pm 0.08 \pm 0.06 \pm 0.04\,\, \mathrm{GeV},
\end{equation}%
and
\begin{equation}
m_{Z_q}=7.11 \pm 0.08 \pm 0.06 \pm 0.01\,\, \mathrm{GeV}.
\end{equation}%
As is seen, obtained in this work predictions are consistent with our results within the errors: The slight
discrepancies between the predictions of this study on the central values of masses of the states under consideration with our
 results can be attributed to the fact that in   Ref.\ \cite{Chen:2013aba} the authors do not take into account some terms in both the light and heavy quark propagators, which are taken into account in the present study. This leads to different working regions of the parameters
$s_0$ and $M^2$ and different situation for the OPE convergence and pole/continuum contribution. 
\begin{table}[tbp]
\begin{tabular}{|c|c|}
\hline\hline
Mass, m.-c. coupling & Results \\ \hline\hline
$m_{Z_q}$ & $(6.97\pm 0.19)~\mathrm{GeV}$ \\
$f_{Z_q} $ & $(0.38 \pm 0.03)\cdot 10^{-2}~\mathrm{GeV}^4 $ \\
$m_{Z_s}$ & $(7.01\pm 0.21)~\mathrm{GeV}$ \\
$f_{Z_s} $ & $(0.41 \pm 0.04)10^{-2}~\mathrm{GeV}^4 $ \\ \hline\hline
\end{tabular}%
\caption{The two-point sum rule prediction for the masses and meson-current
couplings of the $Z_{q}$ and $Z_s$ states.}
\label{tab:Results1}
\end{table}


\section{Widths of the $Z_q \to B_c\protect\pi$, $Z_q \to B_c\protect\eta$ and $Z_s \to B_c\protect\eta$ decay channels}

\label{sec:Width}

In this section we investigate the possible decay channels of the exotic
 $Z_{s(q)}$ states, and calculate the widths of the modes, which are, in
accordance with our results obtained in Sec.\ \ref{sec:Mass}, kinematically
allowed.

It is not dificult to see, that the quark content and mass of the $Z_q$ state
permit its decay to $B_c$ and $\pi$ mesons: The producing of the $B_c$ and
$\eta$ mesons in the decay process is also possible. The tetraquark $Z_s$
can decay to $B_c$ and $\eta$ mesons. At the same time, the modes
$Z_q \to B_c \eta^{\prime}$ and $Z_s \to B_c \eta^{\prime}$ are among
kinematically forbidden decay channels.

We concentrate here on the $Z_s \to B_c \eta$ decay channel. To find its width
we explore the vertex $Z_sB_c\eta$ and calculate the strong coupling
$g_{Z_sB_c \eta}$ using the light cone sum rule method and soft-meson
approximation. To this end, we introduce the following correlation function
\begin{equation}
\Pi(p,q)=i\int d^{4}xe^{ipx}\langle \eta (q)|\mathcal{T}\{J^{B_{c}}(x)J^{%
\dag}(0)\}|0\rangle ,  \label{eq:CorrF3}
\end{equation}
where the interpolating current for the $B_c$ meson is given as
\begin{equation}
J^{B_c}(x)=i\overline{b}_{l}(x)\gamma _{5 }c_{l}(x).  \label{eq:Bcur}
\end{equation}

The correlation function $\Pi (p,q)$ is the basic component of the sum rule
calculations. Expressed in terms of the physical quantities it takes a rather
simple form
\begin{eqnarray}
\Pi ^{\mathrm{Phys}}(p,q) &=&\frac{\langle 0|J^{B_{c}}|B_{c}\left( p\right)
\rangle }{p^{2}-m_{B_{c}}^{2}}\langle B_{c}\left( p\right) \eta
(q)|Z_{s}(p^{\prime })\rangle  \notag \\
&&\times \frac{\langle Z_{s}(p^{\prime })|J^{\dagger}|0\rangle }{p^{\prime
2}-m_{Z}^{2}}+\ldots ,  \label{eq:CorrF4}
\end{eqnarray}%
where $p$, $q$ and $p^{\prime }=p+q$ are the momenta of $B_{c}$, $\eta $%
, and the $Z_{s}$ states, respectively. The first term above is the ground state
contribution, whereas effects of the higher resonances and continuum states are denoted
by the dots.

We define the $B_c$ meson matrix element
\begin{equation}
\langle 0|J^{B_{c}}|B_{c}\left( p\right) \rangle =\frac{%
f_{B_{c}}m_{B_{c}}^{2}}{m_{b}+m_{c}},
\end{equation}
with $m_{B_{c}}$ and $f_{B_{c}}$ being the mass and decay constant of the $%
B_c$ meson, and also the matrix element describing the vertex
\begin{equation}
\langle B_{c}\left( p\right) \eta (q)|Z_{s}(p^{\prime })\rangle
=g_{Z_{s}B_{c}\eta }p\cdot p^{\prime }.  \label{eq:Mel}
\end{equation}
Then the ground state component of the correlation function can be recast into the form:
\begin{equation}
\Pi ^{\mathrm{Phys}}(p,q)=\frac{%
f_{B_{c}}f_{Z}m_{Z}m_{B_{c}}^{2}g_{Z_{s}B_{c}\eta }}{\left( p^{\prime
2}-m_{Z}^{2}\right) \left( p^{2}-m_{B_{c}}^{2}\right) (m_{b}+m_{c})}p\cdot
p^{\prime}.  \label{eq:CorrF5}
\end{equation}%
In the soft-meson limit we apply the restriction $q=0$, which, naturally,
leads to equality  $p=p^{\prime }$ (for details, see Ref.\ \cite{Agaev:2016dev}).
In this approximation the invariant function corresponding to $\Pi ^{\mathrm{Phys}%
}(p,q)$ depends only on the variable $p^{2}$, and is given by the following
expression
\begin{eqnarray}
&&\Pi ^{\mathrm{Phys}}(p^{2})=\frac{%
f_{B_{c}}f_{Z}m_{Z}m_{B_{c}}^{2}g_{Z_{s}B_{c}\eta }}{\left(
p^{2}-m_{Z}^{2}\right) \left( p^{2}-m_{B_{c}}^{2}\right) (m_{b}+m_{c})}m^{2}
\notag \\
&&+\ldots,  \label{eq:CorrF5A}
\end{eqnarray}%
where $m^{2}=\left( m_{Z}^{2}+m_{B_{c}}^{2}\right) /2.$

What is important, now we have to use the one-variable Borel
transformation on $p^{2}$, and apply the operator
\begin{equation}
\left( 1-M^{2}\frac{d}{dM^{2}}\right) M^{2}e^{m^{2}/M^{2}},
\label{eq:softop}
\end{equation}%
to both sides of the sum rule. The last operation is necessary to remove
all unsuppressed contributions emerging in the physical side of the sum
rule due to the soft-meson limit  (see, Ref.\ \cite{Ioffe:1983ju}).

The second side of the sum rule, i.e. QCD expression for
$\Pi ^{\mathrm{QCD}}(p,q)$ is:
\begin{eqnarray}
&&\Pi ^{\mathrm{QCD}}(p,q)=i\int d^{4}xe^{ipx}\left\{ \left[ \gamma _{5}%
\widetilde{S}_{c}^{ib}(x){}\gamma _{5}\right. \right.  \notag \\
&&\left. \times \widetilde{S}_{b}^{bi}(-x){}\gamma _{5}\right] _{\alpha
\beta }\langle \eta (q)|\overline{s}_{\alpha }^{a}s_{\beta }^{a}|0\rangle
\notag \\
&&\left. +\left[ \gamma _{5}\widetilde{S}_{c}^{ib}(x)\gamma _{5}\widetilde{%
S}_{b}^{ai}(-x){}\gamma _{5}\right] _{\alpha \beta }\langle \eta (q)|%
\overline{s}_{\alpha }^{a}s_{\beta }^{b}|0\rangle \right\} .
\end{eqnarray}
Here by $\alpha $ and $\beta $ are the spinor indices.

We proceed by using the expansion
\begin{equation}
\overline{s}_{\alpha }^{a}s_{\beta }^{b}\rightarrow \frac{1}{4}\Gamma
_{\beta \alpha }^{j}\left( \overline{s}^{a}\Gamma ^{j}s^{b}\right) ,
\label{eq:MatEx}
\end{equation}%
where $\Gamma ^{j}$ is the full set of Dirac matrixes, and performing the
summation over color indices.

Calculation of the traces over spinor indices, and integration of the obtained
integrals in accordance with procedures reported in Ref.\ \cite%
{Agaev:2016dev} enable us to extract the imaginary part of the correlation function $%
\Pi ^{\mathrm{QCD}}(p,q)$. As a result, we find not only the spectral density,
but also determine local matrix elements of the $\eta$ meson that form it. Our analysis
proves that in the soft-meson limit only the  local twist-3 matrix element
$\langle \eta (q)|\overline{s}i\gamma _{5}s|0\rangle$  survives and contributes
to the spectral density $\rho_{\eta}^{s}(s)$ corresponding to the $Z_{s}B_c\eta$ vertex.
Within the same approximation the strong couplings of the vertices
$Z_{q}B_c\eta$ and $Z_{q}B_c\pi$ are determined by the matrix elements
$\langle \eta (q)|\overline{q}i\gamma _{5}q|0\rangle$ and $\langle \pi (q)|\overline{q}i\gamma _{5}q|0\rangle$, respectively.

Situation with the pion is clear: its matrix element is known, and was used in our previous
works to explore decays of other tetraquarks. But the matrix elements of the eta mesons
deserve more detailed analysis, which is connected with mixing phenomena in the
$\eta-\eta^{\prime} $ system.

The $\eta-\eta^{\prime}$ mixing and $U(1)$ axial anomaly are problems which decisively
affect  physics of the eta mesons.  The $\eta-\eta^{\prime}$ mixing can be described
using either the singlet-octet basis of the flavor group $SU_f(3)$, or the
quark-flavor basis. The latter is founded on the $\bar s s$ and $(\bar u u +
\bar d d)/\sqrt{2}$ as the basic states, and is convenient to describe the
mixing phenomena of the $\eta-\eta^{\prime}$ system, including  mixing of the physical
states, decay constants and higher twist distribution amplitudes
[Ref.\ \cite{Feldmann:1998vh,Beneke:2002jn,Agaev:2014wna,Agaev:2015faa}].

In the present work we follow this approach and utilize the quark-flavor
mixing scheme in our calculations. Then the twist-3 matrix elements of interest are given as
\begin{eqnarray}
&&2m_{q}\langle \eta (q)|\overline{q}i\gamma _{5}q|0\rangle =\frac{h_{\eta
}^{q}}{\sqrt{2}}, \\
&&2m_{s}\langle \eta (q)|\overline{s}i\gamma _{5}s|0\rangle =h_{\eta }^{s},
\end{eqnarray}
where the parameters $h_{\eta}^{s(q)}$ are defined by the equalities
\begin{align}
h_{\eta}^{s (q)} &= m_{\eta}^2 f_{\eta}^{s(q)} - A_{\eta}\,,  \notag \\
A_{\eta} &= \langle 0 | \frac{\alpha_s}{4\pi}G^a_{\mu\nu}\widetilde
G^{a,\mu\nu}|\eta(p)\rangle,  \label{eq:aM}
\end{align}
and $A_{\eta}$ is the matrix element appeared due to the $U(1)$ anomaly .

In Refs.\ \cite{Beneke:2002jn,Agaev:2014wna,Agaev:2015faa} it was assumed
that the parameters $h_{\eta}^{s(q)}$ obey the same mixing scheme as the
decay constants of the eta mesons, and hence the following equality holds:
\begin{equation}
\begin{pmatrix}
h_{\eta }^{q} & h_{\eta }^{s} \\
h_{\eta ^{\prime }}^{q} & h_{\eta ^{\prime }}^{s}%
\end{pmatrix}%
=%
\begin{pmatrix}
\cos \varphi & -\sin \varphi \\
\sin \varphi & \cos \varphi%
\end{pmatrix}%
\begin{pmatrix}
h_{q} & 0 \\
0 & h_{s}%
\end{pmatrix}
.
\end{equation}
Here $\varphi$ is the mixing angle in the quark-flavor scheme, $h_s$ and
$h_q$ are input parameters extracted from analysis of
the experimental data:
\begin{align}
h_q =& (0.0016 \pm 0.004)\,\, \mathrm{GeV}^3,  \notag \\
h_s =& (0.087\pm 0.006)\,\, \mathrm{GeV}^3,  \notag \\
\varphi =& 39.3^{\circ}\pm 1.0^\circ.  \label{FKSvalues}
\end{align}

The details about the local matrix elements of the eta mesons presented above,
is sufficient to calculate the spectral densities under investigation. We find:

\begin{equation}
\rho _{\eta }^{s}(s) =\frac{h_{\eta }^{s}}{48m_{s}}L(s) ,
\end{equation}

for the  $Z_{s}B_c\eta$ vertex,
\begin{equation}
\rho _{\eta }^{q}(s) =\frac{h_{\eta }^{q}}{48\sqrt{
2}m_{q}}L(s) , 
\end{equation}
for the  $Z_{q}B_c\eta$ vertex,
and
\begin{equation}
\rho _{\pi }(s)=\frac{f_{\pi} m_{\pi }^2}{24\sqrt{%
2}m_q}L(s)
\end{equation}
for the  $Z_{q}B_c\pi$ vertex, where the "universal" function $L(s)$ has the form
\begin{eqnarray}
L(s)&=&\frac{1}{\pi^2 s^2}\left[ s^{2}+s\left( m_{b}^{2}+6m_{b}m_{c}+m_{c}^{2}\right) -2(m_{b}^{2}-m_{c}^{2})^2\right] 
\notag \\
&\times & \sqrt{\left(
s+m_{b}^{2}-m_{c}^{2}\right) ^{2}-4m_{b}^{2}s}+\frac{1}{3}\int\limits_{0}^{1}  \frac{dz}{j^2 z^2}
\notag \\
&\times & \left\lbrace \langle\alpha_{s}\frac{G^{2}}{\pi}\rangle\left[ s(m_b^2j^3+m_bm_cjz-m_c^2z^3) \delta^{(2)}\left( s-\Phi\right)\right. \right. 
\notag \\
&+&\left.\left.  2 \left(m_b^2j^3-m_c^2z^3+m_bm_c(1+3jz) \right) \delta^{(1)}\left( s-\Phi\right)
\right]
\right. 
\notag \\
&+&\left.\langle g^3 G^3\rangle \frac{1}{5\times 2^6\pi^2j^{3}z^{3}}\right. 
\notag \\
&\times&\left. \left\lbrace 12j^2z^2 \left[  3m_bm_c(1+5jz(1+jz))\right. \right. \right. 
\notag \\
&+&\left.\left.\left.   3m_b^2j^5-z\left(3m_c^2z^4+sj(1+jz(7+11jz)) \right)\right] 
\right. \right. 
\notag \\
&\times&\left.\left. 
 \delta^{(2)}\left( s-\Phi\right)-2jz\left[m_c^3z^5(4m_b-7m_c)
 \right. \right. \right. 
\notag \\
&+&\left.\left. \left. 
  2s^2j^3z^3(2+7jz)+m_b^2j^5\left(7m_b^2-4m_bm_c 
  \right. \right. \right. \right. 
\notag \\
&+&\left.\left.\left. \left. 
  9s(1-2z)z \right)+9m_csjz^2\left( m_cz^3(2z-1)
  \right. \right. \right. \right. 
\notag \\
&-&\left.\left.\left. \left. 
  2m_bj(1+3jz)\right)  \right] \delta^{(3)}\left( s-\Phi\right)\right. \right. 
  \notag \\
&+&\left.  \left. 
  \left[ 2m_b^5m_cj^5-2m_c^5m_bz^5-s^3j^5z^5+6s^2j^3z^3
  \right. \right. \right. 
  \notag \\
&\times&\left. \left.  \left. 
  \left( m_b^2j^3+m_bm_cjz-m_c^2z^3\right)+sjz\left( 4m_b^3m_cj^4z
   \right. \right. \right. \right. 
  \notag \\
&-&\left. \left. \left. \left. 
  7m_b^4j^5-4m_c^3m_bjz^4+7m_c^4z^5\right)  \right]  \delta^{(4)}\left( s-\Phi\right)\right\rbrace \right. 
   \notag \\
   &+&\left. \langle\alpha_{s}\frac{G^{2}}{\pi}\rangle^2 \frac{ m_bm_c}{3^3\times 2}
 \left[-6jz \delta^{(3)}\left( s-\Phi\right)
  \right. \right. 
  \notag \\
&+&\left.\left. 
  2\left(m_bm_c-s(1+3jz) \right) \delta^{(4)}\left( s-\Phi\right)
  \right. \right. 
  \notag \\
&+&\left.\left. 
  s\left(m_bm_c-sjz \right)\delta^{(5)}\left( s-\Phi\right)  \right] 
\right\rbrace ,
\end{eqnarray}%
where
\begin{eqnarray}
\Phi &=& \frac{m_b^2j-m_c^2z}{jz},
\notag \\
j&=&z-1.
\end{eqnarray}

The final sum rule to evaluate the strong coupling reads
\begin{eqnarray}
&&g_{Z_{s}B_{c}\eta }=\frac{(m_{b}+m_{c})}{%
f_{B_{c}}f_{Z}m_{Z}m_{B_{c}}^{2}m^{2}}\left( 1-M^{2}\frac{d}{dM^{2}}%
\right) M^{2}  \notag \\
&&\times \int_{\mathcal{M}^2}^{s_{0}}dse^{(m^{2}-s)/M^{2}}\rho_{\eta}^{s}(s).  \label{eq:SRules}
\end{eqnarray}

The similar expressions are valid for the remaining two couplings  $g_{Z_{q}B_{c}\eta }$ and $g_{Z_{q}B_{c}\pi}$, as well.

In order to get the width of the decay $Z_{s}\to B_{c}\eta$ we adapt to this case
the expression derived in Ref.\ \cite{Agaev:2016ijz}, which takes the form
\begin{eqnarray}
&&\Gamma \left( Z_{s}\to B_{c}\eta\right) =\frac{%
g_{Z_{s}B_{c}\eta }^{2}m_{B_{c}}^{2}}{24\pi }\lambda \left( m_{Z},\
m_{B_{c}},m_{\eta }\right)   \notag \\
&&\times \left[ 1+\frac{\lambda ^{2}\left( m_{Z_{s}},\ m_{B_{c}},m_{\eta
}\right) }{m_{B_{c}}^{2}}\right] ,  \label{eq:DW}
\end{eqnarray}%
where
\begin{equation*}
\lambda (a,\ b,\ c)=\frac{\sqrt{a^{4}+b^{4}+c^{4}-2\left(
a^{2}b^{2}+a^{2}c^{2}+b^{2}c^{2}\right) }}{2a}.
\end{equation*}%

Parameters required for numerical computations of the decay widths
are listed in Table\ \ref{tab:Param}. Apart from the standard information it
contains also the decay constant $f_{B_c}$ of the $B_c$ meson, for which we
utilize its value derived in the context of the sum rule method in Ref.\ \cite{Baker:2013mwa}.

The analysis carried out in accordance with traditional requirements
of the sum rule calculations enable us to fix the working windows for
the parameters $s_0$ and $M^2$ in this section. Our  analyses show that the same regions for the $M^2$  and  $s_0$ as the mass sum rules in the previous section lead to a better convergence of OPE and a nice pole contribution for the strong coupling constants under consideration.  The perturbative-nonperturbative comparison, convergence  of nonperturbative series and pole/total ratio as an example  for $ Z_q B_c \pi $ vertex are depicted in Figs. \ref{fig:pert_nonpertg}-\ref{fig:PCg}. From these figures we see that the perturbative contribution exceeds the nonperturbative one considerably and the OPE nicely converges. We also get a nice pole contribution of about $ 70\% $. Similar results are obtained for other vertices. 
\begin{widetext}

\begin{figure}[h!]
\begin{center}
\includegraphics[totalheight=6cm,width=8cm]{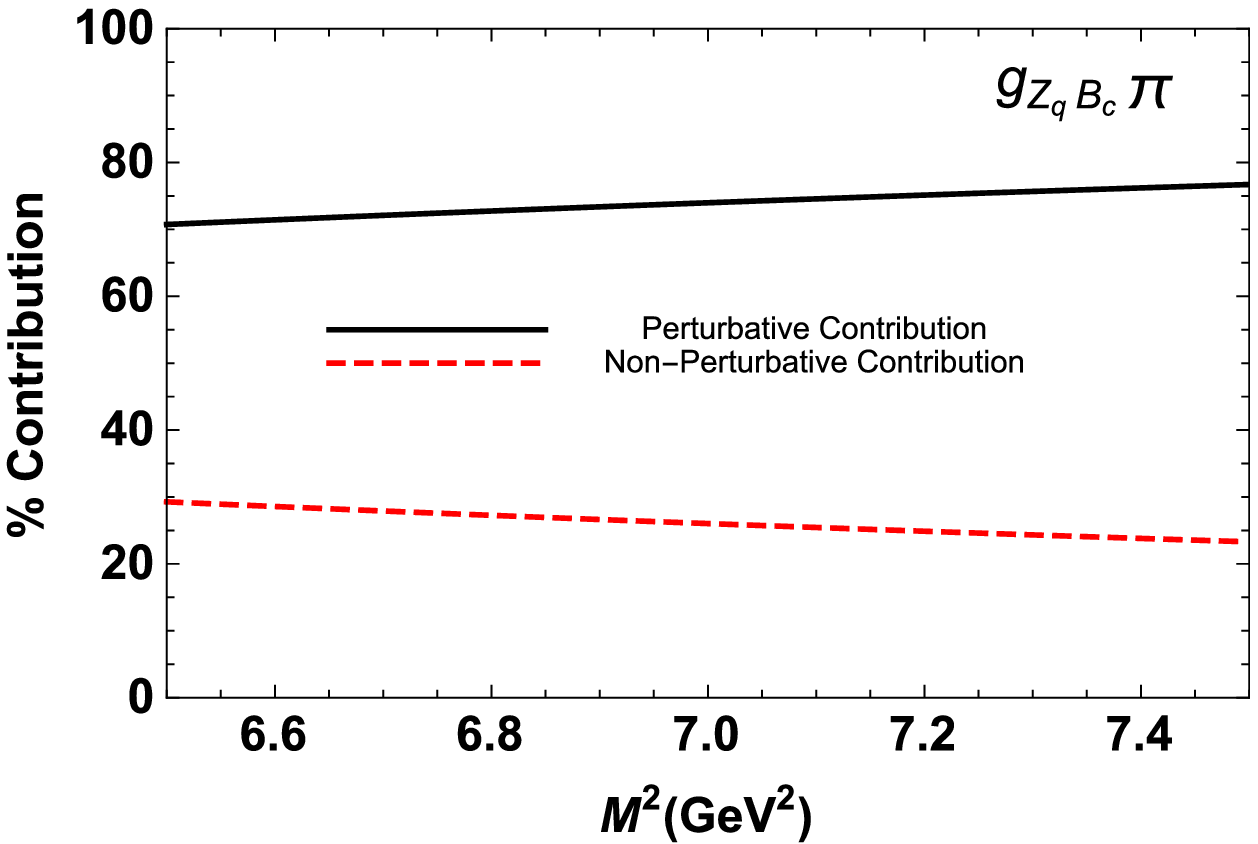}\,\,
\includegraphics[totalheight=6cm,width=8cm]{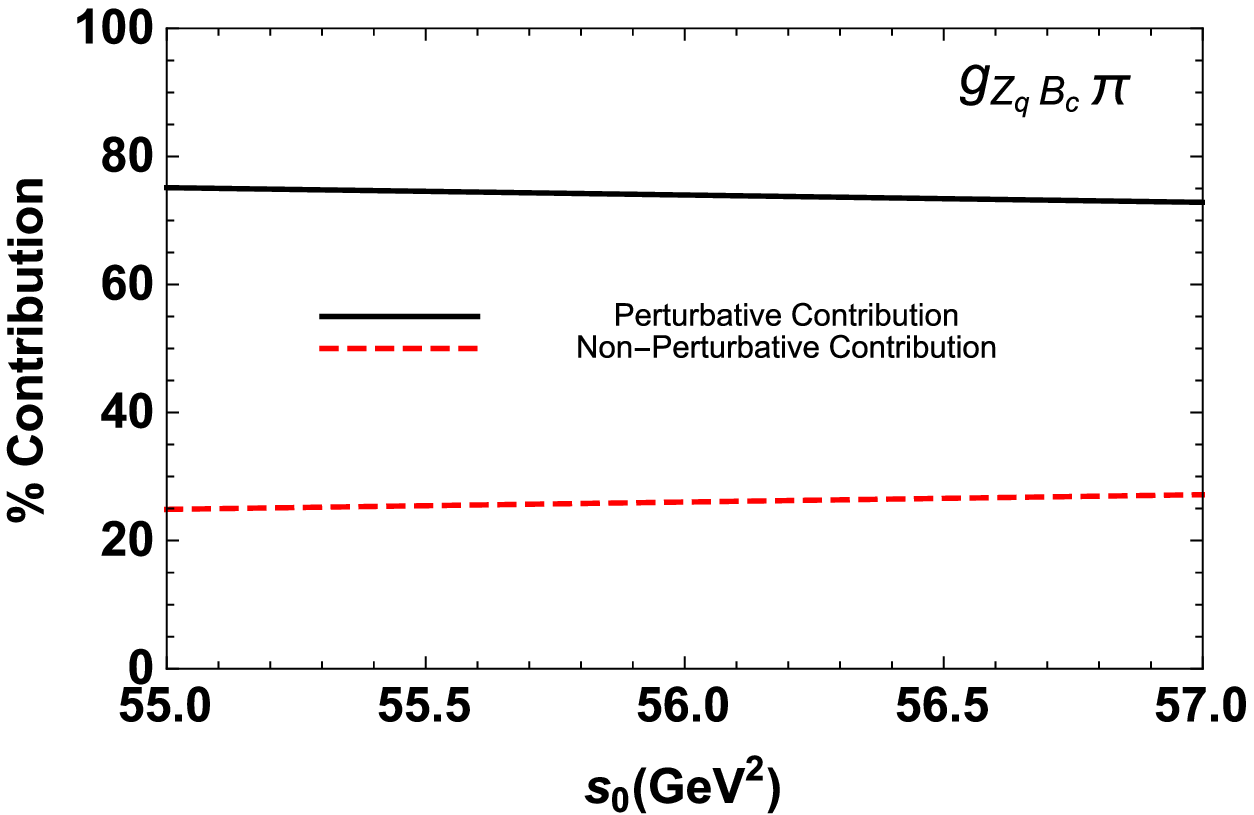}
\end{center}
\caption{\textbf{Left:} Comparison of  the perturbative and nonperturbative contributions to the  $ Z_q B_c \pi $ vertex with respect to  $M^2$ at average value of $ s_0 $. \textbf{Right:}
 The same as left panel but in terms of $ s_0 $ at average value of the   Borel
parameter $M^2$. } \label{fig:pert_nonpertg}
\end{figure}
\begin{figure}[h!]
\begin{center}
\includegraphics[totalheight=6cm,width=8cm]{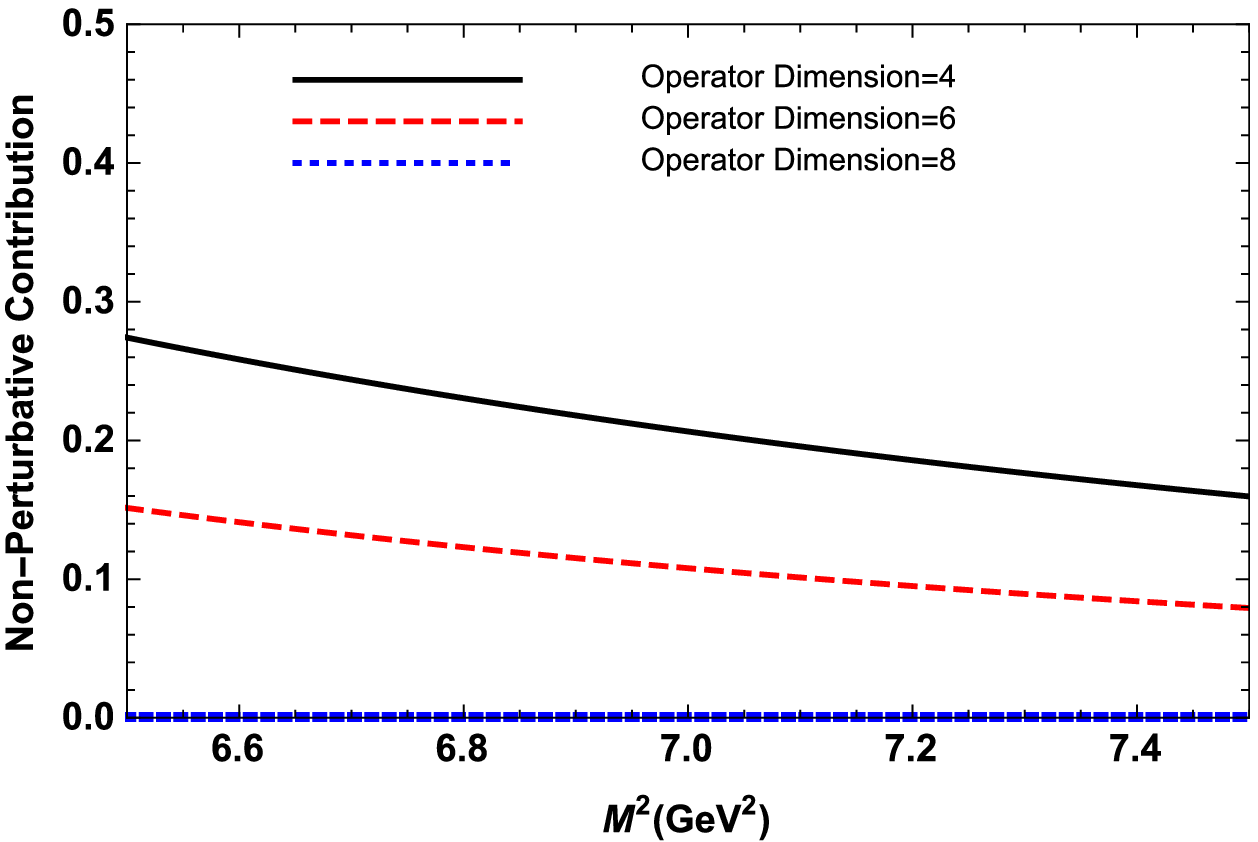}\,\,
\includegraphics[totalheight=6cm,width=8cm]{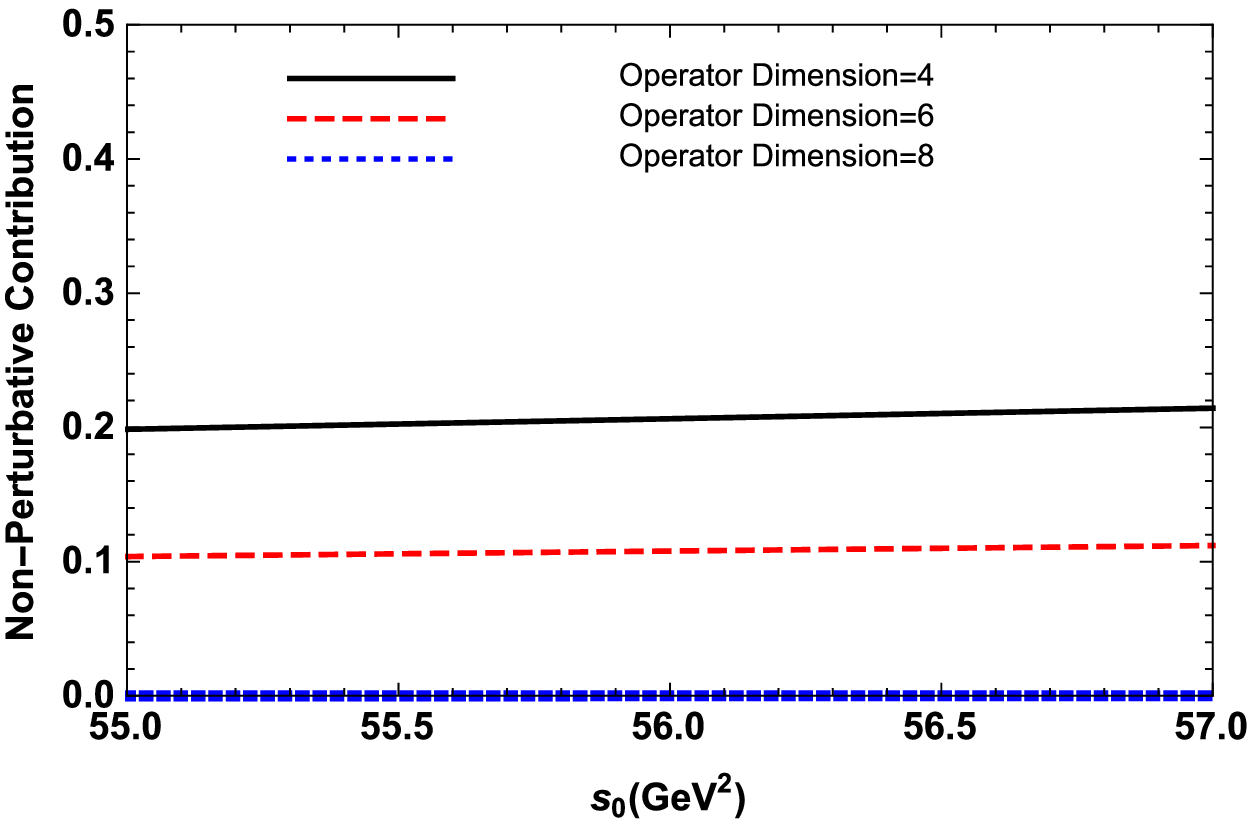}
\end{center}
\caption{\textbf{Left:} Contribution of different  nonperturbative operators to the  $ Z_q B_c \pi $ vertex with respect to  $M^2$ at average value of $ s_0 $. \textbf{Right:}
 The same as left panel but in terms of $ s_0 $ at average value of the   Borel
parameter $M^2$. } \label{fig:nonpertg}
\end{figure}
\begin{figure}[h!]
\begin{center}
\includegraphics[totalheight=6cm,width=8cm]{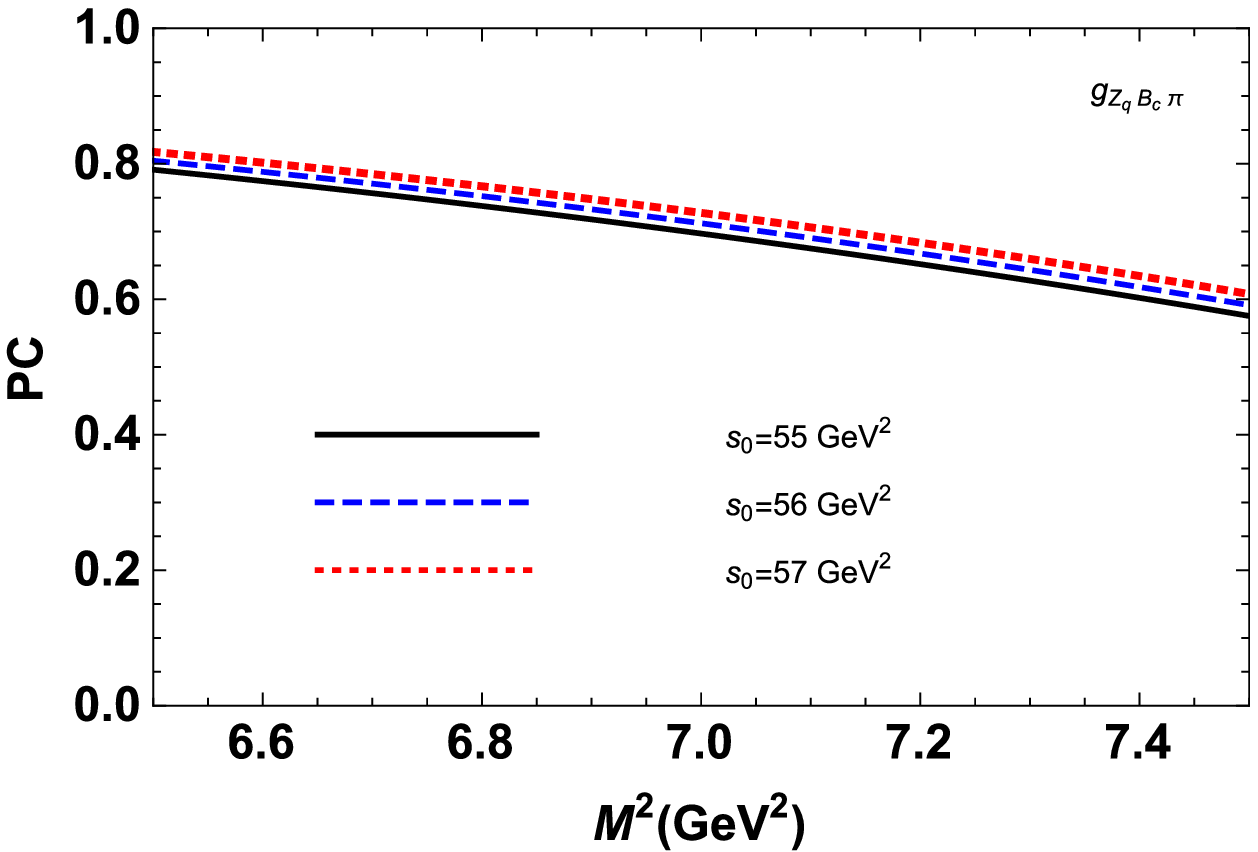}\,\,
\includegraphics[totalheight=6cm,width=8cm]{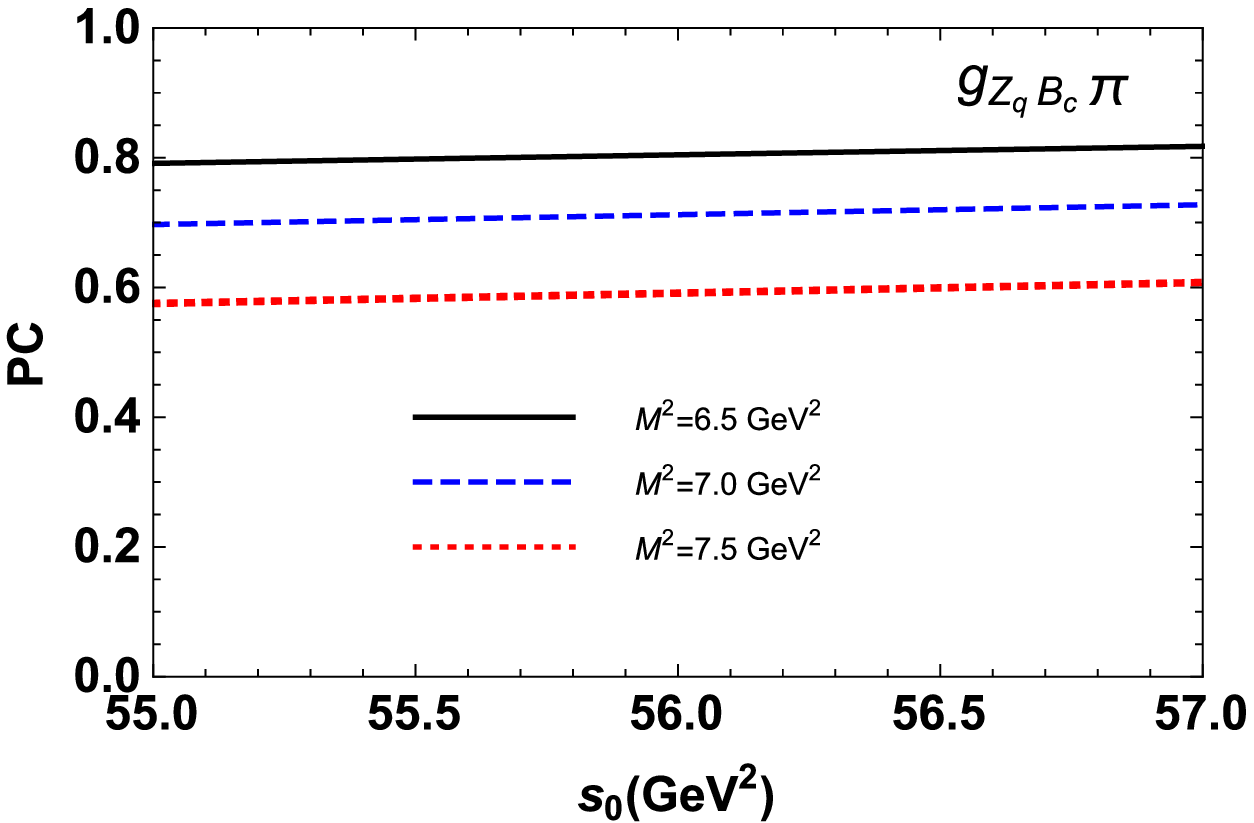}
\end{center}
\caption{\textbf{Left:} Pole/total contribution of  $ Z_q B_c \pi $ vertex with respect to  $M^2$ at different fixed values of $ s_0 $. \textbf{Right:}
 The same as left panel but in terms of $ s_0 $ at different fixed values of the   Borel
parameter $M^2$. } \label{fig:PCg}
\end{figure}
\begin{figure}[h!]
\begin{center}
\includegraphics[totalheight=6cm,width=8cm]{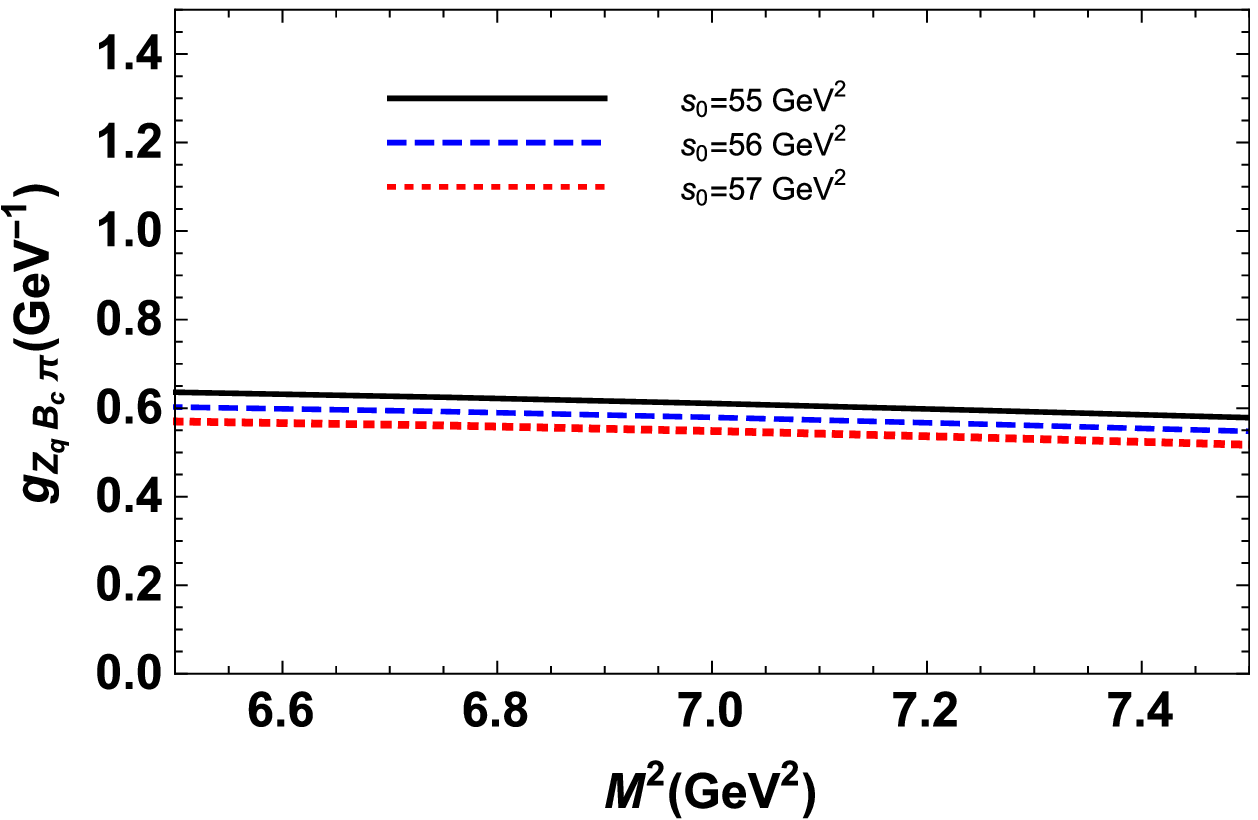}
\includegraphics[totalheight=6cm,width=8cm]{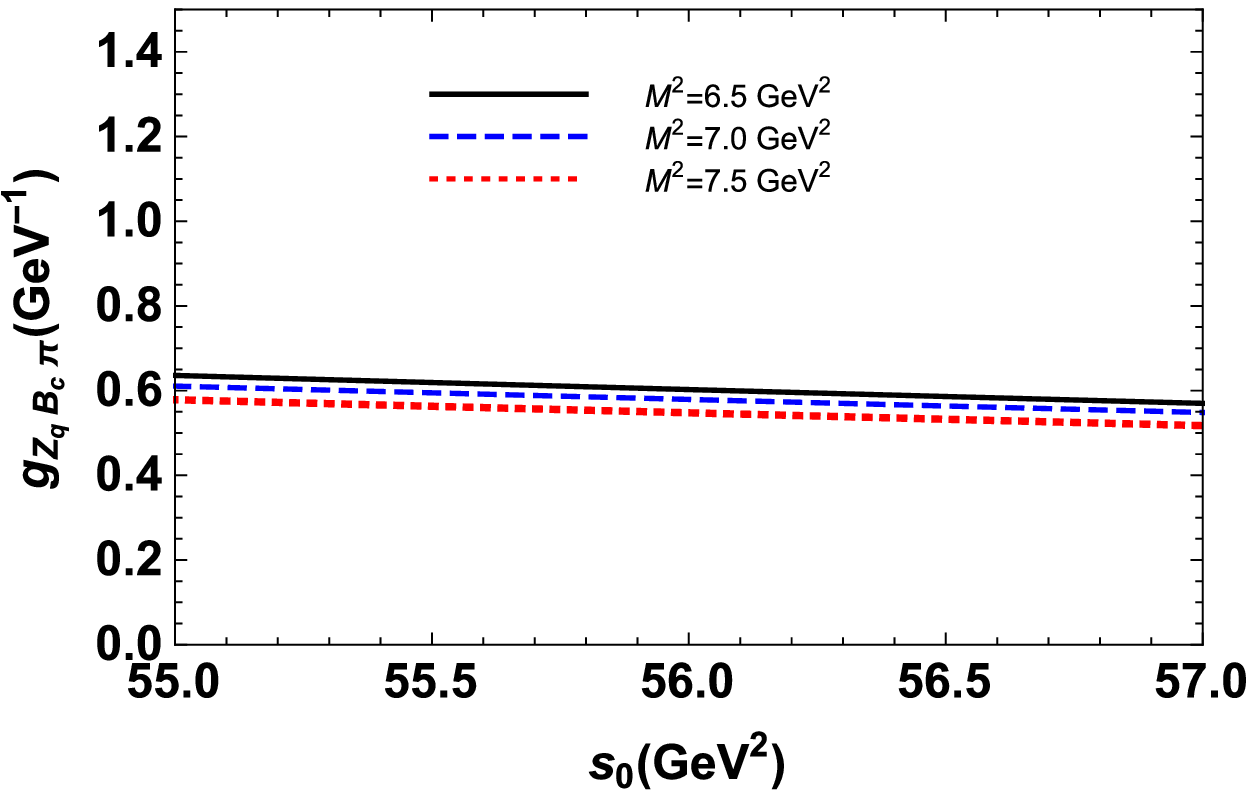}
\end{center}
\caption{\textbf{Left:} The  coupling constant $g_{Z_{q}B_c \pi}$ as a function of the Borel parameter $
M^2$ at various values of $s_0$. \textbf{Right:}
 The  coupling constant $g_{Z_{q}B_c \pi}$ as a function of threshold $
s_0$ at various values of $M^2$.} \label{fig:coplingZqpi}
\end{figure}
\begin{figure}[h!]
\begin{center}
\includegraphics[totalheight=6cm,width=8cm]{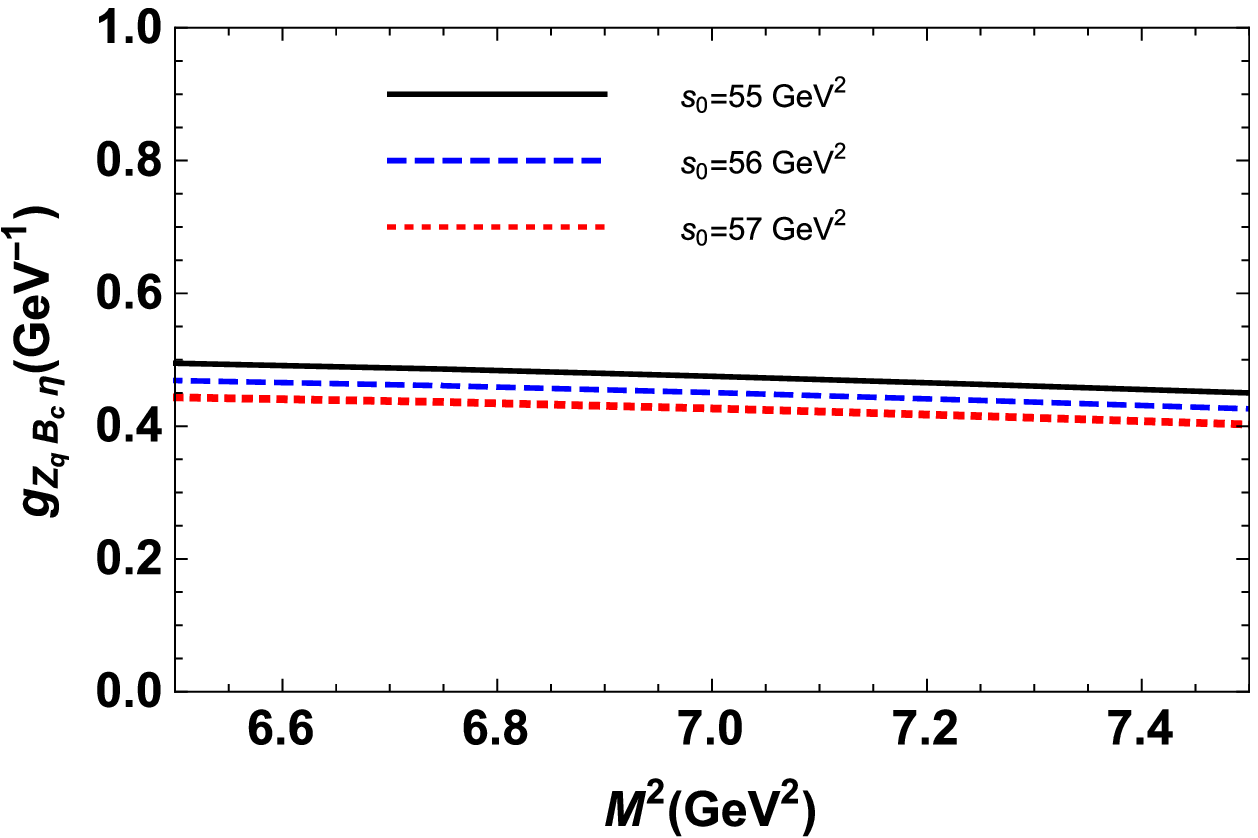}
\includegraphics[totalheight=6cm,width=8cm]{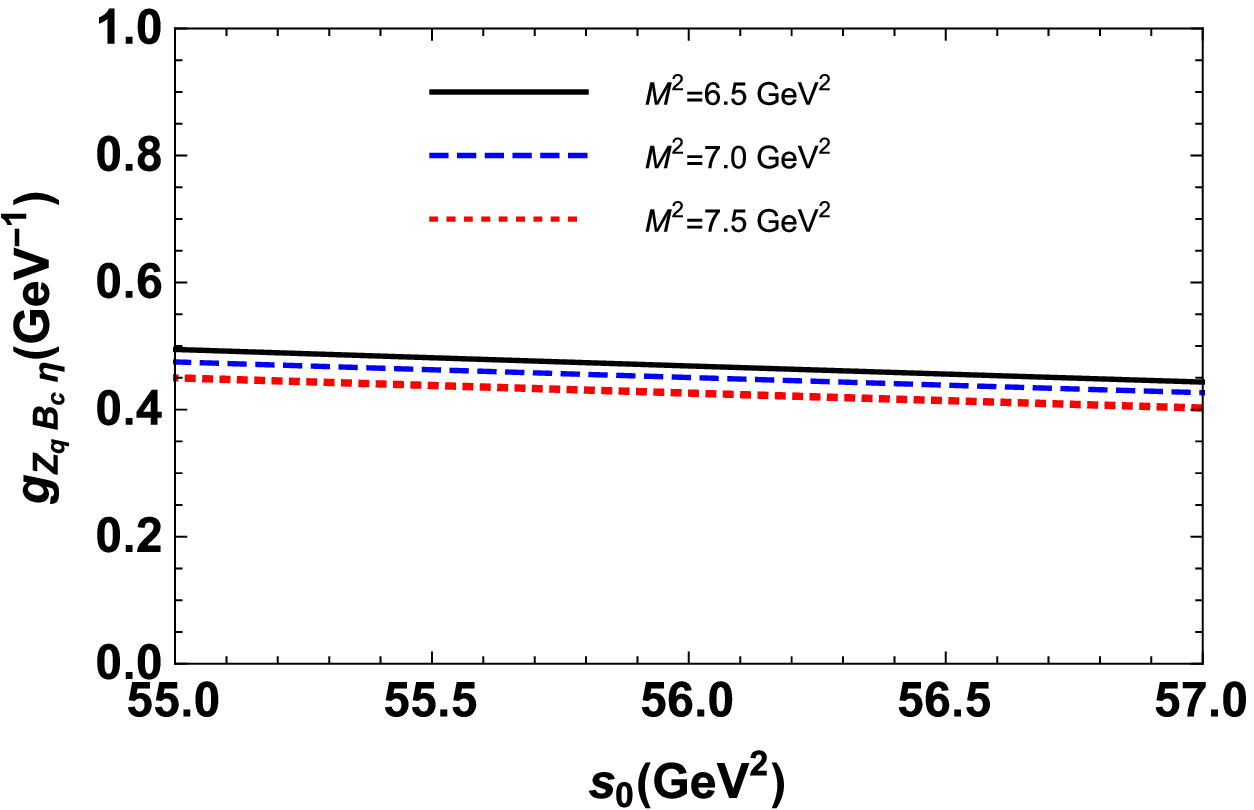}
\end{center}
\caption{\textbf{Left:} The  coupling constant $g_{Z_{q}B_c \eta}$ as a function of the Borel parameter $
M^2$ at various values of $s_0$. \textbf{Right:}
 The  coupling constant $g_{Z_{q}B_c \eta}$ as a function of threshold $
s_0$ at various values of $M^2$.} \label{fig:coplingZqeta}
\end{figure}
\begin{figure}[h!]
\begin{center}
\includegraphics[totalheight=6cm,width=8cm]{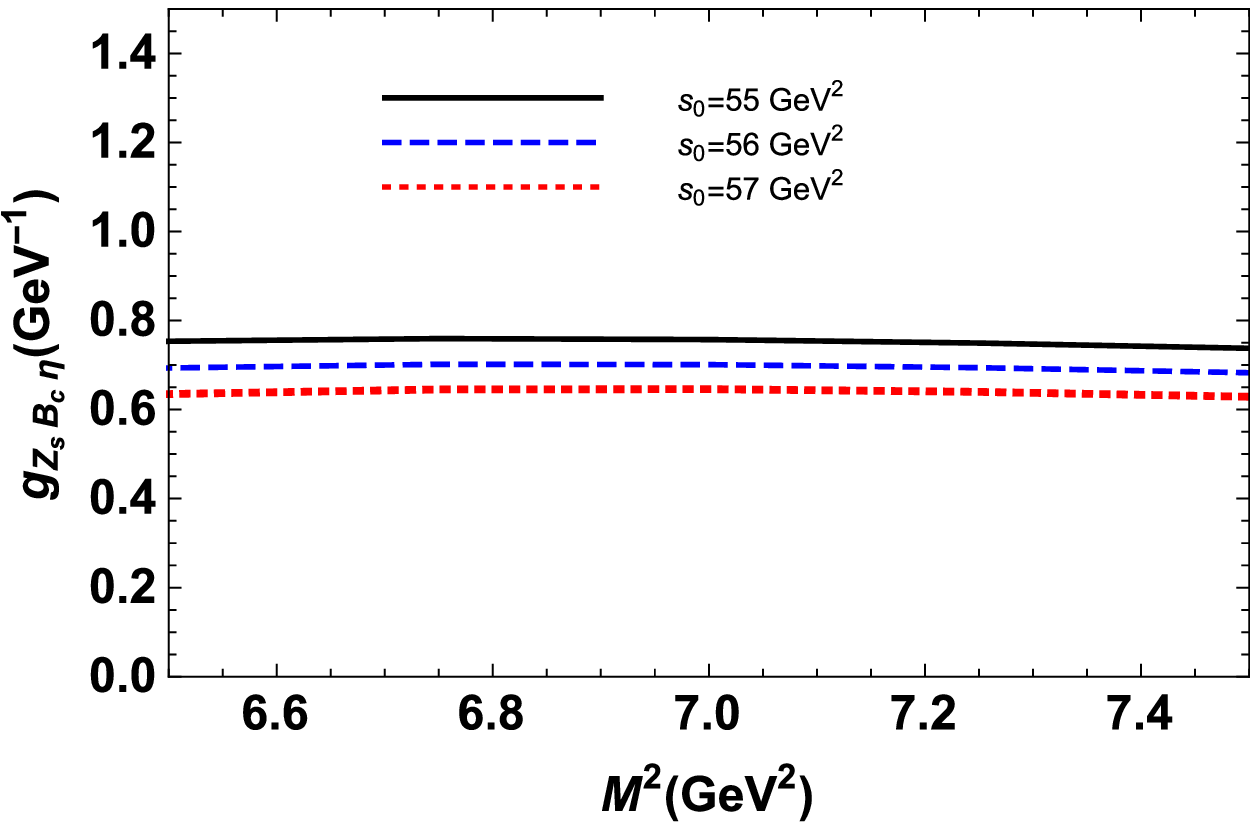}
\includegraphics[totalheight=6cm,width=8cm]{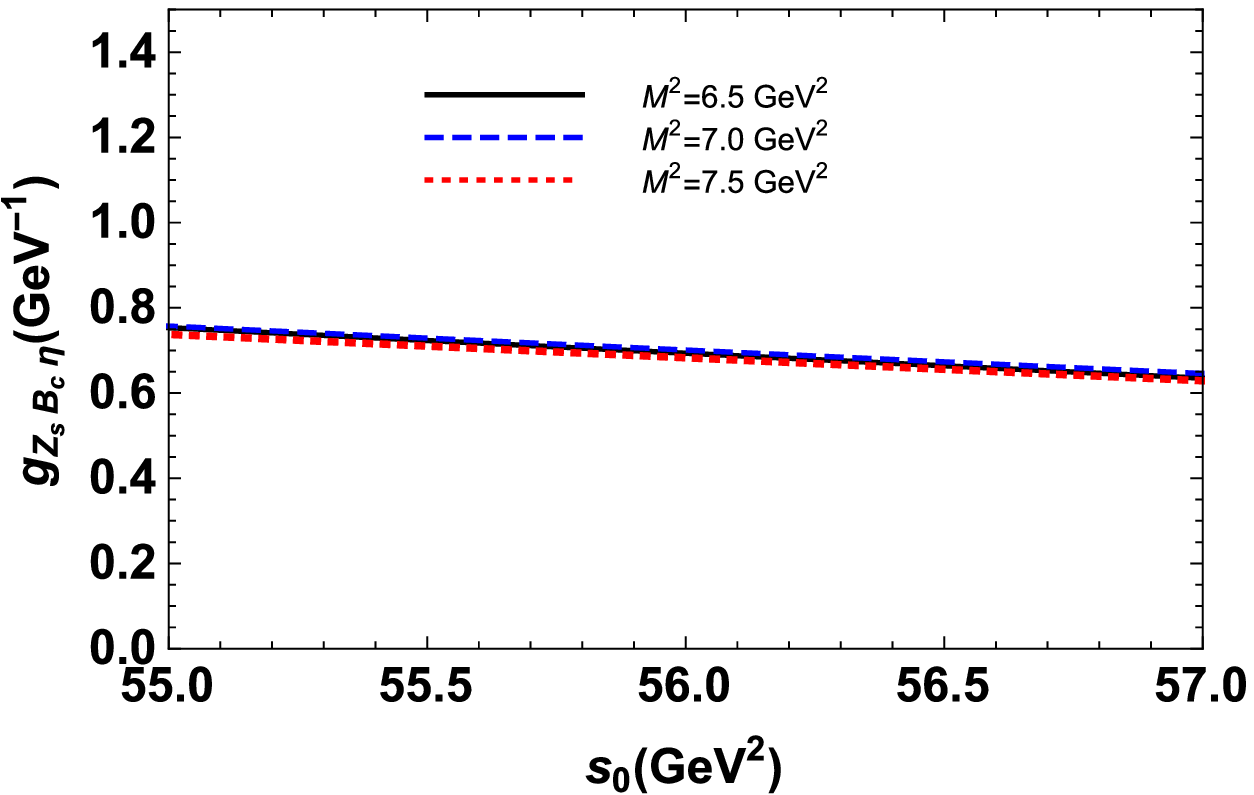}
\end{center}
\caption{\textbf{Left:} The  coupling constant $g_{Z_{s}B_c \eta}$ as a function of the Borel parameter $
M^2$ at various values of $s_0$. \textbf{Right:}
 The  coupling constant $g_{Z_{s}B_c \eta}$ as a function of threshold $
s_0$ at various values of $M^2$.} \label{fig:coplingZseta}
\end{figure}
\end{widetext}


Depicted in Figs.\ \ref{fig:coplingZqpi}-\ref{fig:coplingZseta} output of numerical calculations
demonstrates the dependence  of the strong coupling constants, $g_{Z_{q}B_c \pi}$,  $g_{Z_{q}B_c \eta}$ and $g_{Z_{s}B_c \eta}$ on $M^2$ and $s_0$,
which demonstrate good instabilities of the couplings with respect to auxiliary parameters. 

The strong couplings and decay widths of the exploring processes
are collected in Table\ \ref{tab:Results2}. The obtained results
are typical for the decays of tetraquark states. One of their notable features
is the difference between $\Gamma(Z_q \to B_c \pi)$ and $\Gamma(Z_q \to B_c \eta)$. In fact,
the $Z_q$ state may interact with the pion and $\eta$ meson through its $\bar q q$ component.
But the spectral density of the vertex $Z_q B_c \eta$ is proportional to $h_{\eta}^q$, which
numerically is considerably smaller than $f_{\pi}m_{\pi}^2$ entering into $\rho_{\pi}(s)$. The
reason is a reducing effect of the axial anomaly explicit from Eq.\ (\ref{eq:aM}).
\begin{table}[tbp]
\begin{tabular}{|c|c|}
\hline\hline
Strong couplings, Widths & Predictions \\ \hline\hline
$g_{Z_{q}B_c \pi}$ & $(0.57 \pm 0.21) $ $\mathrm{GeV}^{-1}$ \\
$g_{Z_{q}B_c \eta}$ & $(0.45 \pm 0.17) $ $\mathrm{GeV}^{-1}$ \\
$g_{Z_{s}B_c \eta}$ & $(0.69 \pm 0.26) $ $\mathrm{GeV}^{-1}$ \\
$\Gamma(Z_q \to B_c \pi)$ & $(111 \pm 49)$ $\mathrm{MeV}$ \\
$\Gamma(Z_q \to B_c \eta)$ & $(43 \pm 19)$ $\mathrm{MeV}$ \\
$\Gamma(Z_s \to B_c \eta)$ & $(112 \pm 51)$ $\mathrm{MeV}$ \\ \hline\hline
\end{tabular}%
\caption{The strong couplings and decay widths of the $Z_{q}$ and $Z_s$
exotic particles obtained within the soft-meson approximation. }
\label{tab:Results2}
\end{table}

Investigation of the open charm-bottom tetraquarks performed in the present work within the
diquark-antidiquark picture led to quite interesting predictions. Theoretical exploration of these
states using alternative pictures for their internal organization, as well as  experimental
studies may shed light not only on their parameters but also on properties of the conventional
particles.

\section*{ACKNOWLEDGEMENTS}

Work of K.~A. was  financed by T\"{U}B\.{I}TAK under the Grant no: 115F183.

\appendix*

\section{ The  spectral densities for $ Z_q $ state}

\renewcommand{\theequation}{\Alph{section}.\arabic{equation}} \label{sec:App}

Here we present the results obtained for the two-point spectral
density corresponding to the   $ Z_q $ state. We get
\begin{equation}
\rho ^{\mathrm{QCD}}(s)=\rho ^{\mathrm{pert.}}(s)+\sum_{k=3}^{8}\rho _{k}(s),
\label{eq:A1}
\end{equation}%
where by $\rho _{k}(s)$ we
denote the nonperturbative contributions to $\rho ^{\mathrm{QCD}}(s)$. The
explicit expressions for $\rho ^{\mathrm{pert}}(s)$ and $\rho _{k}(s)$ are
obtained in terms of  the integrals of the Feynman parameters $z$ and $ w $ as

\begin{widetext}
\begin{eqnarray} 
\label{eq:A2}
\rho ^{\mathrm{pert}}(s)&=&\frac{1}{2^8 \pi^6}\int\limits_{0}^{1} dz\int\limits_{0}^{1-z} dw\frac{wz}{ht^8}\left(m_b^2tw+m_c^2tz-shwz \right)^{2} 
\nonumber\\
&\times& 
\left[ z^2 \left(6m_c^2shtw-7s^2h^2w^2-m_c^4t^2 \right)+2m_b^2twz\left(3shw-m_c^2t\right)-m_b^4t^2w^2  \right] \Theta\left[L\left(s,z,w\right)  \right],
\notag \\
\rho _{\mathrm{3}}(s)&=&\frac{\langle
\overline{q}q\rangle}{2^3 \pi^4}\int\limits_{0}^{1} dz\int\limits_{0}^{1-z} dw \frac{\left(m_b w+m_c z \right) }{t^5}\left( 2shwz-m_c^2rz-m_b^2tw\right) \left(m_b^2tw+m_c^2rz-shwz \right) \Theta\left[L\left(s,z,w\right)  \right],
\notag \\
\rho _{\mathrm{4}}(s)&=&-\frac{\langle
\alpha_{s}\frac{G^{2}}{\pi}\rangle}{3\times 2^9 \pi^4}\int\limits_{0}^{1} dz\int\limits_{0}^{1-z} dw \frac{wz}{ht^6}\left\lbrace z^2\left[ 30h^3s^2w^3-4sm_c^2hrw\left(9pw+9wz+4z^2 \right)\right. \right. 
\notag \\
&+&\left. \left. m_c^4 r^2\left(9pw+9wz+8z^2 \right)  \right]+2m_b^2twz\left[m_c^2t\left( 9wz+13w^2-9w+4z^2\right)-2shw^2\left(13w+9z-9 \right)   \right]\right. 
\notag \\
&+&\left. m_b^4t^2w^3\left( 17w+9z-9\right)    \right\rbrace \Theta\left[L\left(s,z,w\right)  \right],
\notag \\
\rho _{\mathrm{5}}(s)&=&\frac{m_0^2\langle
\overline{q}q\rangle}{2^{5}\times \pi^{4}}\int\limits_{0}^{1} dz\int\limits_{0}^{1-z} dw \frac{h\left( m_bw+m_cz\right) }{t^{4}}\left(2m_b^{2}tw+2m_c^{2}rz-3shwz \right) \Theta\left[L\left(s,z,w\right)  \right],
\notag \\
\rho _{\mathrm{6}}(s)&=&\frac{\langle
g_{s}^{3}G^{3}\rangle}{5\times 3\times 2^{12}\pi^6}\int\limits_{0}^{1} dz\int\limits_{0}^{1-z} dw\frac{wz}{hf^2t^7}\left\lbrace 28m_b^2w^7f^5+zw^6f^5\left(32m_b^2+10m_c^2-21s\right)+w^5f^4z^2\right. 
\notag \\
&\times & \left. 
\left[ 2m_c^2f+s\left( 11w-32\right)+m_b^2\left( 30w+2\right)   \right] +2w^4z^3\left[3m_b^2j+2s-17m_b^2w-m_c^2f^4(3w-4)\right.\right.
\notag \\
&+&\left. \left.  m_b^2w^2(38-42w+23w^2-5w^3)+2s(11-54w+86w^2-59w^3+15w^4) \right]   -2w^3z^4
\right.
\notag \\
&\times &\left.   \left[ j^2(m_b^2(19w-4)-3m_c^2-2s)-2sw(w^2-2)(4-7w+4w^2)+m_b^2w^2(10w^3-43w^2+73w-61)
\right.\right. 
\notag \\
&+ &\left. \left. 
m_c^2w(23-62w+78w^2-47w^3+11w^4)\right] +2w^2z^5\left[m_b^2j+m_c^2j^2(1-7w)+3m_b^2j^2w(4w-1)
\right.\right. 
\notag \\
&+ &\left. \left.
2sj^2(31w-8)+m_b^2w^3(56w-14w^2-83)+m_c^2w^2(46-70w+49w^2-13w^3)+2sw^2(5w+13w^2
\right.\right. 
\notag \\
&- &\left. \left.
6w^3-33) \right]+z^6w\left[j^3(21s-32m_c^2+2m_b^2(6w-5))+j^2w(12m_b^2w(f+w)+s(25+22w)
\right.\right. 
\notag \\
&+&\left. \left.
m_c^2(78w-90))+w^3(2m_c^2(f-1)(11f-5)+2m_b^2w(16w-51)+s(163w-19w^2-350)) \right]
\right. 
\notag \\
&-&\left. 
z^7\left[ s(228w^4-75w^{5})+26m_b^2w^{5}+20m_c^{2}w^{5}-2j^{2}w^{2}\left(3m_c^{2}(f-10)-5m_b^{2}w+3s(3+w) \right)
\right. \right. 
\notag \\
&+&\left. \left. 
2j^{3}\left( 2m_c^{2}(12w-7)+3sw+m_b^{2}w(5w-2)\right)  \right] +z^{8}\left[j^{2}w^{2}(8m_c^{2}+15s)+2w^{4}(19s+5m_c^{2})
\right. \right. 
\notag \\
&+&\left. \left.
j^{3}\left(m_c^{2}(4-10w)-2m_b^{2}w+3sw \right)  \right]-2j^{3}z^{9}m_c^{2}   
\right\rbrace \Theta\left[L\left(s,z,w\right)  \right]
\notag \\
&+&
\frac{
g_{s}^{2}\langle
\overline{q}q\rangle^{2}}{3^2\times 2\pi^{4}}\int\limits_{0}^{1} dz\int\limits_{0}^{1-z} dw  \frac{h^2wz}{t^5}\left(2shwz-m_b^2tw-m_c^2tz \right)  \Theta\left[L\left(s,z,w\right)  \right]
\notag \\
&+&\frac{\langle
\overline{q}q\rangle^{2}}{6\pi^{2}}\frac{m_bm_c}{s}\sqrt{(s+m_b^{2}-m_c^{2})^{2}-4sm_b^{2}},
\notag \\
\rho _{\mathrm{7}}(s)&=&\frac{\langle
\alpha_{s}\frac{G^{2}}{\pi}\rangle \langle
\overline{q}q\rangle}{3^2\times 2^{5}\pi^2}\left\lbrace \int\limits_{0}^{1} dz\int\limits_{0}^{1-z} dw \frac{1}{\left[  w^2+j(w+z)\right]^4 }\left\lbrace 4m_bw^3(z^2+zf-2wf)+m_cz\left[  fwz(3-2z-6w)\right. \right. \right. 
\notag \\
&+&\left. \left.\left.  z^3(w-8z+8)-3f^2w^2 \right]   \right\rbrace  \Theta\left[L\left(s,z,w\right)  \right]+\frac{m_b+m_c}{s^2}\left[ (m_b-m_c)^2-s\right]\sqrt{(m_b^2-m_c^2+s)^2-4sm_b^2}  \right\rbrace ,
\notag \\
\rho _{\mathrm{8}}(s)&=&-\frac{\langle
\alpha_{s}\frac{G^{2}}{\pi}\rangle^{2} }{3^4\times 2^9 \pi^{2}}\int\limits_{0}^{1} dz\int\limits_{0}^{1-z} dw \frac{m_b^2m_c^2wz}{h^4t^2f} \left\lbrace hfwz \left[10 \delta^{(1)}\left(s-\Delta \right) +11\delta^{(2)}\left(s- \Delta \right) \right] +2s^2t^2\delta^{(3)}\left(s-\Delta \right) \right\rbrace, 
\end{eqnarray}
where we omitted to show the terms proportional to the $ m_q $ in order to avoid from  very lengthy 
expressions. Here,
\begin{eqnarray}
L\left(s,z,w\right) &=& -\frac{f\left[j(w+z)(m_b^2w+m_c^2z)+w(m_b^2w^2-shz+m_c^2wz) \right] }{\left( w^2+j(w+z)\right)^2 },
\notag \\
\delta^{(n)}\left(s- \Delta \right)&=&\left( \frac{d}{ds}\right) ^{n}\left(s- \Delta \right),
\notag \\
\Delta &=& \frac{t(m_b^2w+m_c^2z)}{hwz},
\notag \\
t &=& w^2+(w+z)(z-1),
\notag \\
r &=& z^2+(w+z)(w-1),
\notag \\
h &=& w+z-1,
\notag \\
f &=& w-1,
\notag \\
j &=& z-1,
\end{eqnarray}
and $  \Theta\left[...\right] $ is the usual unit-step function. 

\end{widetext}

\end{document}